\documentclass[pdflatex,sn-basic]{sn-jnl}   

\AtBeginDocument{\setlength{\evensidemargin}{\oddsidemargin}}

\newif\ifanon
\anonfalse

\usepackage[T1]{fontenc}      
\usepackage[utf8]{inputenc}
\usepackage{amsmath,amssymb}  
\usepackage{array}
\usepackage{booktabs}
\usepackage{adjustbox}       
\usepackage{flafter}         
\usepackage{tikz}            
\usetikzlibrary{positioning,arrows.meta}
\usepackage{graphicx}
\usepackage{enumitem}
\usepackage{listings}
\usepackage{url}
\usepackage{hyperref}
\usepackage[capitalise,noabbrev]{cleveref}
\usepackage{etoolbox}
\crefname{appendix}{Appendix}{Appendices}
\AtBeginEnvironment{appendices}{\crefalias{section}{appendix}}

\newcommand{\cat}[1]{\texttt{\small #1}}
\newcommand{\code}[1]{\texttt{\small #1}}
\newcommand{\idpattern}{\code{\textasciicircum{}[a-z][a-z0-9\_]\{2,30\}\$}}
\newcommand{\idpatternloose}{\code{\textasciicircum{}[a-zA-Z0-9][a-zA-Z0-9\_\textbackslash-]\{2,30\}\$}}

\newcommand{\pn}[1]{\ifcsname pn@#1\endcsname\csname pn@#1\endcsname\else\PackageError{numbers}{Unknown number key #1}{Regenerate paper_numbers.json}\fi}
\expandafter\def\csname pn@ablation.aggressive.attempts\endcsname{189}
\expandafter\def\csname pn@ablation.aggressive.length_bound_incl_concealed\endcsname{30.2\%}
\expandafter\def\csname pn@ablation.baseline.attempts\endcsname{1{,}066}
\expandafter\def\csname pn@ablation.baseline.failure_rate\endcsname{64.9\%}
\expandafter\def\csname pn@ablation.baseline.length_bound\endcsname{14.4\%}
\expandafter\def\csname pn@ablation.baseline.length_bound_incl_concealed\endcsname{37.8\%}
\expandafter\def\csname pn@ablation.baseline.markdown_wrap\endcsname{49.0\%}
\expandafter\def\csname pn@ablation.few_shot.attempts\endcsname{176}
\expandafter\def\csname pn@ablation.few_shot.failure_rate\endcsname{50.0\%}
\expandafter\def\csname pn@ablation.few_shot.length_bound\endcsname{22.7\%}
\expandafter\def\csname pn@ablation.few_shot.length_bound_incl_concealed\endcsname{36.9\%}
\expandafter\def\csname pn@ablation.few_shot.markdown_wrap\endcsname{25.0\%}
\expandafter\def\csname pn@c.claude-haiku-4-5@medium@financebench.fence_compound_share\endcsname{71.2\%}
\expandafter\def\csname pn@c.claude-haiku-4-5@medium@financebench.questions\endcsname{18}
\expandafter\def\csname pn@c.claude-haiku-4-5@medium@hotpotqa.attempts_per_run\endcsname{7.96}
\expandafter\def\csname pn@c.claude-haiku-4-5@medium@hotpotqa.fence_compound_share\endcsname{51.0\%}
\expandafter\def\csname pn@c.claude-opus-4-8@medium@hotpotqa.attempts_per_run\endcsname{3.04}
\expandafter\def\csname pn@c.claude-opus-4-8@medium@hotpotqa.failures\endcsname{2}
\expandafter\def\csname pn@c.claude-sonnet-5@medium@financebench.failures\endcsname{24}
\expandafter\def\csname pn@c.claude-sonnet-5@medium@financebench.rate\endcsname{21.1\%}
\expandafter\def\csname pn@c.claude-sonnet-5@medium@financebench.share.length_bound\endcsname{41.7\%}
\expandafter\def\csname pn@c.claude-sonnet-5@medium@financebench.share.length_bound.ci\endcsname{[25.0, 61.1]}
\expandafter\def\csname pn@c.claude-sonnet-5@medium@financebench.share.markdown_wrap\endcsname{25.0\%}
\expandafter\def\csname pn@c.claude-sonnet-5@medium@financebench.share.regex_pattern\endcsname{33.3\%}
\expandafter\def\csname pn@c.claude-sonnet-5@medium@hotpotqa.dominant_share\endcsname{81.4\%}
\expandafter\def\csname pn@c.claude-sonnet-5@medium@hotpotqa.failures\endcsname{43}
\expandafter\def\csname pn@c.claude-sonnet-5@medium@hotpotqa.fence_compound_share\endcsname{8.6\%}
\expandafter\def\csname pn@c.claude-sonnet-5@medium@hotpotqa.rate\endcsname{21.6\%}
\expandafter\def\csname pn@calib.field.answer.max\endcsname{500}
\expandafter\def\csname pn@calib.field.answer.mean_overrun\endcsname{15.2\%}
\expandafter\def\csname pn@calib.field.answer.min\endcsname{417}
\expandafter\def\csname pn@calib.field.answer.n\endcsname{9}
\expandafter\def\csname pn@calib.field.coverage_notes.max\endcsname{315}
\expandafter\def\csname pn@calib.field.coverage_notes.max_overrun\endcsname{26.0\%}
\expandafter\def\csname pn@calib.field.coverage_notes.mean_overrun\endcsname{9.0\%}
\expandafter\def\csname pn@calib.field.coverage_notes.min\endcsname{251}
\expandafter\def\csname pn@calib.field.coverage_notes.n\endcsname{8}
\expandafter\def\csname pn@calib.field.fact.max\endcsname{480}
\expandafter\def\csname pn@calib.field.fact.n\endcsname{1}
\expandafter\def\csname pn@calib.medium.failures\endcsname{28}
\expandafter\def\csname pn@calib.medium.length_bound\endcsname{18}
\expandafter\def\csname pn@calib.medium.rate\endcsname{15.7\%}
\expandafter\def\csname pn@calib.medium_v2.failures\endcsname{24}
\expandafter\def\csname pn@calib.medium_v2.length_bound\endcsname{0}
\expandafter\def\csname pn@calib.medium_v2.rate\endcsname{7.4\%}
\expandafter\def\csname pn@calib.medium_v2.regex_pattern\endcsname{14}
\expandafter\def\csname pn@cf.invalid_to_valid.length_bound\endcsname{18}
\expandafter\def\csname pn@cf.n\endcsname{178}
\expandafter\def\csname pn@cf.samev2.failures\endcsname{10}
\expandafter\def\csname pn@collection.date_max\endcsname{2026-08-28}
\expandafter\def\csname pn@collection.date_min\endcsname{2026-08-09}
\expandafter\def\csname pn@conceal.exh.fenced\endcsname{47}
\expandafter\def\csname pn@conceal.exh.len_feedback_once\endcsname{23}
\expandafter\def\csname pn@conceal.exh.len_feedback_twice\endcsname{1}
\expandafter\def\csname pn@conceal.exh.n\endcsname{24}
\expandafter\def\csname pn@conceal.exh.persist\endcsname{24}
\expandafter\def\csname pn@conceal.exh.same_field\endcsname{24}
\expandafter\def\csname pn@crossfield.not_reached\endcsname{1{,}122}
\expandafter\def\csname pn@crossfield.reached\endcsname{2{,}875}
\expandafter\def\csname pn@crossfield.validated\endcsname{2{,}871}
\expandafter\def\csname pn@crossfield.violations\endcsname{0}
\expandafter\def\csname pn@divergence.gap_pp\endcsname{0.4}
\expandafter\def\csname pn@divergence.groups_qualifying\endcsname{6}
\expandafter\def\csname pn@divergence.groups_total\endcsname{12}
\expandafter\def\csname pn@divergence.pairs\endcsname{15}
\expandafter\def\csname pn@divergence.runner_up_category_pp\endcsname{72.5}
\expandafter\def\csname pn@divergence.top.0.pp\endcsname{72.9}
\expandafter\def\csname pn@divergence.top.n_tied\endcsname{4}
\expandafter\def\csname pn@divergence.unrestricted.max.n_tied\endcsname{2}
\expandafter\def\csname pn@divergence.unrestricted.max.pp\endcsname{97.5}
\expandafter\def\csname pn@excluded_pilot.attempts\endcsname{34}
\expandafter\def\csname pn@excluded_variant.attempts\endcsname{365}
\expandafter\def\csname pn@excluded_variant.runs\endcsname{60}
\expandafter\def\csname pn@fence.reported.concealed.length_bound\endcsname{327}
\expandafter\def\csname pn@fence.reported.concealed.missing_field\endcsname{1}
\expandafter\def\csname pn@fence.reported.concealed.regex_pattern\endcsname{26}
\expandafter\def\csname pn@fence.reported.n\endcsname{681}
\expandafter\def\csname pn@fence.reported.pure_share\endcsname{48.0\%}
\expandafter\def\csname pn@full.attempts\endcsname{4{,}396}
\expandafter\def\csname pn@full.cells\endcsname{20}
\expandafter\def\csname pn@full.failures\endcsname{1{,}326}
\expandafter\def\csname pn@g.claude-haiku-4-5@medium.att1.cat.markdown_wrap\endcsname{165}
\expandafter\def\csname pn@g.claude-haiku-4-5@medium.att1.n\endcsname{475}
\expandafter\def\csname pn@g.claude-haiku-4-5@medium.att1.valid\endcsname{240}
\expandafter\def\csname pn@g.claude-haiku-4-5@medium.attempts_per_completed_run\endcsname{9.36}
\expandafter\def\csname pn@g.claude-haiku-4-5@medium.completed\endcsname{142}
\expandafter\def\csname pn@g.claude-haiku-4-5@medium.completion_rate\endcsname{85.5\%}
\expandafter\def\csname pn@g.claude-haiku-4-5@medium.dominant_share\endcsname{72.9\%}
\expandafter\def\csname pn@g.claude-haiku-4-5@medium.exhausted\endcsname{24}
\expandafter\def\csname pn@g.claude-haiku-4-5@medium.exhausted.node.gather\endcsname{23}
\expandafter\def\csname pn@g.claude-haiku-4-5@medium.fa.failed\endcsname{475}
\expandafter\def\csname pn@g.claude-haiku-4-5@medium.fa.reached\endcsname{475}
\expandafter\def\csname pn@g.claude-haiku-4-5@medium.fa.retry_success\endcsname{451}
\expandafter\def\csname pn@g.claude-haiku-4-5@medium.failures\endcsname{878}
\expandafter\def\csname pn@g.claude-haiku-4-5@medium.fence_compound_ci\endcsname{[49.9, 59.9]}
\expandafter\def\csname pn@g.claude-haiku-4-5@medium.fence_compound_share\endcsname{54.7\%}
\expandafter\def\csname pn@g.claude-haiku-4-5@medium.fence_n\endcsname{640}
\expandafter\def\csname pn@g.claude-haiku-4-5@medium.rate\endcsname{66.1\%}
\expandafter\def\csname pn@g.claude-haiku-4-5@medium.residual_rate\endcsname{44.2\%}
\expandafter\def\csname pn@g.claude-haiku-4-5@medium.runs\endcsname{166}
\expandafter\def\csname pn@g.claude-opus-4-8@medium.dominant_share\endcsname{76.9\%}
\expandafter\def\csname pn@g.claude-opus-4-8@medium.failures\endcsname{13}
\expandafter\def\csname pn@g.claude-opus-4-8@medium.share.length_bound.ci\endcsname{[53.8, 100.0]}
\expandafter\def\csname pn@g.claude-sonnet-4-6@medium.rate\endcsname{15.7\%}
\expandafter\def\csname pn@g.claude-sonnet-5@medium.dominant_share\endcsname{61.2\%}
\expandafter\def\csname pn@g.claude-sonnet-5@medium.fence_compound\endcsname{4}
\expandafter\def\csname pn@g.claude-sonnet-5@medium.fence_compound_ci\endcsname{[2.1, 19.4]}
\expandafter\def\csname pn@g.claude-sonnet-5@medium.fence_n\endcsname{41}
\expandafter\def\csname pn@g.claude-sonnet-5@medium.residual_rate\endcsname{9.6\%}
\expandafter\def\csname pn@g.gpt-4o-mini@medium.completed\endcsname{69}
\expandafter\def\csname pn@g.gpt-4o-mini@medium.dominant_share\endcsname{66.7\%}
\expandafter\def\csname pn@g.gpt-4o-mini@medium.rate\endcsname{13.6\%}
\expandafter\def\csname pn@g.gpt-4o-mini@medium.runs\endcsname{70}
\expandafter\def\csname pn@g.gpt-4o@medium.dominant_share\endcsname{72.5\%}
\expandafter\def\csname pn@g.gpt-4o@medium.rate\endcsname{16.0\%}
\expandafter\def\csname pn@heldout.agree\endcsname{69}
\expandafter\def\csname pn@heldout.ambiguous\endcsname{20}
\expandafter\def\csname pn@heldout.cat.length_bound.agree\endcsname{12}
\expandafter\def\csname pn@heldout.cat.length_bound.n\endcsname{30}
\expandafter\def\csname pn@heldout.cat.markdown_wrap.agree\endcsname{34}
\expandafter\def\csname pn@heldout.cat.markdown_wrap.n\endcsname{40}
\expandafter\def\csname pn@heldout.cat.regex_pattern.agree\endcsname{23}
\expandafter\def\csname pn@heldout.cat.regex_pattern.n\endcsname{30}
\expandafter\def\csname pn@heldout.ci\endcsname{[59.4, 77.2]}
\expandafter\def\csname pn@heldout.dis.fence_ambiguous\endcsname{2}
\expandafter\def\csname pn@heldout.dis.fence_underlying\endcsname{4}
\expandafter\def\csname pn@heldout.dis.identifier_as_enum\endcsname{2}
\expandafter\def\csname pn@heldout.dis.identifier_pattern\endcsname{7}
\expandafter\def\csname pn@heldout.dis.length_unverified\endcsname{18}
\expandafter\def\csname pn@heldout.disagree\endcsname{31}
\expandafter\def\csname pn@heldout.n\endcsname{100}
\expandafter\def\csname pn@heldout.nonambiguous.rate\endcsname{86.2\%}
\expandafter\def\csname pn@heldout.rate\endcsname{69.0\%}
\expandafter\def\csname pn@heldout.weighted_rate\endcsname{73.4\%}
\expandafter\def\csname pn@hidden.exhausted_attempts\endcsname{3}
\expandafter\def\csname pn@hidden.gpt4omini.share_without\endcsname{63.3\%}
\expandafter\def\csname pn@hidden.n\endcsname{4}
\expandafter\def\csname pn@m.claude-opus-4-8.attempts\endcsname{265}
\expandafter\def\csname pn@m.claude-sonnet-4-6.attempts\endcsname{502}
\expandafter\def\csname pn@matched.hotpotqa_medium.shared_questions\endcsname{50}
\expandafter\def\csname pn@meta.bootstrap_reps\endcsname{2{,}000}
\expandafter\def\csname pn@meta.db_sha256_short\endcsname{d0f406f00882}
\expandafter\def\csname pn@meta.git_head_short\endcsname{31f41e168b}
\expandafter\def\csname pn@modellevel.violations\endcsname{4}
\expandafter\def\csname pn@pert.medium.gpt-5.6-sol.failures\endcsname{3}
\expandafter\def\csname pn@pert.medium.gpt-5.6-terra.failures\endcsname{6}
\expandafter\def\csname pn@pert.medium.identifiers\endcsname{8}
\expandafter\def\csname pn@pert.medium.leading_digit\endcsname{4}
\expandafter\def\csname pn@pert.medium.length_overflow\endcsname{4}
\expandafter\def\csname pn@pert.medium_v3_loose_regex.gpt-5.6-sol.failures\endcsname{5}
\expandafter\def\csname pn@pert.medium_v3_loose_regex.gpt-5.6-terra.failures\endcsname{4}
\expandafter\def\csname pn@pert.medium_v3_loose_regex.identifiers\endcsname{11}
\expandafter\def\csname pn@pert.medium_v3_loose_regex.leading_digit\endcsname{0}
\expandafter\def\csname pn@pert.medium_v3_loose_regex.length_overflow\endcsname{10}
\expandafter\def\csname pn@pert.pooled.failures\endcsname{18}
\expandafter\def\csname pn@pert.pooled.identifiers\endcsname{19}
\expandafter\def\csname pn@pert.pooled.regex_failures\endcsname{17}
\expandafter\def\csname pn@pooled.reported.att0.n\endcsname{2{,}896}
\expandafter\def\csname pn@pooled.reported.att0.valid\endcsname{2{,}192}
\expandafter\def\csname pn@pooled.reported.attempts_per_completed_run\endcsname{4.22}
\expandafter\def\csname pn@pooled.reported.ci\endcsname{[26.3, 30.1]}
\expandafter\def\csname pn@pooled.reported.completed\endcsname{948}
\expandafter\def\csname pn@pooled.reported.fa.failed\endcsname{704}
\expandafter\def\csname pn@pooled.reported.fa.reached\endcsname{2{,}896}
\expandafter\def\csname pn@pooled.reported.fa.retry_success\endcsname{679}
\expandafter\def\csname pn@pooled.reported.fa.valid\endcsname{2{,}192}
\expandafter\def\csname pn@pooled.reported.fa.valid_after_fence_rate\endcsname{85.8\%}
\expandafter\def\csname pn@pooled.reported.fa.valid_rate\endcsname{75.7\%}
\expandafter\def\csname pn@pooled.reported.rate\endcsname{28.2\%}
\expandafter\def\csname pn@pooled.reported.residual_ci\endcsname{[17.6, 22.5]}
\expandafter\def\csname pn@pooled.reported.residual_rate\endcsname{20.0\%}
\expandafter\def\csname pn@pooled.reported.runs\endcsname{973}
\expandafter\def\csname pn@reported.attempts\endcsname{3{,}997}
\expandafter\def\csname pn@reported.cat.length_bound\endcsname{263}
\expandafter\def\csname pn@reported.categories_observed\endcsname{7}
\expandafter\def\csname pn@reported.cells\endcsname{18}
\expandafter\def\csname pn@reported.exhaiku.rate\endcsname{9.3\%}
\expandafter\def\csname pn@reported.failures\endcsname{1{,}126}
\expandafter\def\csname pn@reported.length_bound_incl_concealed\endcsname{590}
\expandafter\def\csname pn@reported.medium.macro_rate\endcsname{16.7\%}
\expandafter\def\csname pn@retrygap.cause.length\endcsname{328}
\expandafter\def\csname pn@retrygap.concealed_msgs\endcsname{354}
\expandafter\def\csname pn@retrygap.concealed_share\endcsname{32.2\%}
\expandafter\def\csname pn@retrygap.feedback_msgs\endcsname{1{,}101}
\expandafter\def\csname pn@retrygap.model.claude-haiku-4-5.concealed_msgs\endcsname{350}
\expandafter\def\csname pn@retrygap.runnodes\endcsname{218}
\expandafter\def\csname pn@retrygap.runnodes_exhausted\endcsname{24}
\expandafter\def\csname pn@robust.cell.attempts\endcsname{1{,}066}
\expandafter\def\csname pn@robust.cell.failures\endcsname{692}
\expandafter\def\csname pn@robust.cell.questions\endcsname{50}
\expandafter\def\csname pn@robust.cell.rate\endcsname{64.9\%}
\expandafter\def\csname pn@robust.cell.runs\endcsname{134}
\expandafter\def\csname pn@robust.one_run.attempts\endcsname{404}
\expandafter\def\csname pn@robust.one_run.failures\endcsname{260}
\expandafter\def\csname pn@robust.one_run.rate\endcsname{64.4\%}
\expandafter\def\csname pn@robust.within_cell.max_abs_delta_pp\endcsname{1.3}
\expandafter\def\csname pn@screen.fence_any.markdown_wrap\endcsname{789}
\expandafter\def\csname pn@screen.fence_any.matches\endcsname{790}
\expandafter\def\csname pn@screen.fence_any.prose_preamble\endcsname{1}
\expandafter\def\csname pn@screen.nonjson_opener.matches\endcsname{9}
\expandafter\def\csname pn@screen.nonjson_opener.prose_preamble\endcsname{9}
\expandafter\def\csname pn@screen.properties.matches\endcsname{42}
\expandafter\def\csname pn@screen.properties.schema_mirroring\endcsname{42}
\expandafter\def\csname pn@validation.blind_agree\endcsname{140}
\expandafter\def\csname pn@validation.blind_ci\endcsname{[93.1, 98.9]}
\expandafter\def\csname pn@validation.blind_rate\endcsname{97.2\%}
\expandafter\def\csname pn@validation.n\endcsname{144}
\expandafter\def\csname pn@versions.gpt-4o-mini.attempts\endcsname{242}
\expandafter\def\csname pn@versions.gpt-4o-mini.precapture\endcsname{69}
\expandafter\def\csname pn@versions.gpt-4o.attempts\endcsname{250}
\expandafter\def\csname pn@versions.gpt-4o.precapture\endcsname{72}
\expandafter\def\csname pn@zero_length.model\endcsname{claude-sonnet-5}
\expandafter\def\csname pn@zero_length.node\endcsname{gather}
\expandafter\def\csname pn@zero_length.output_tokens\endcsname{4{,}096}

\newcommand{\rqanswer}[2]{\par\medskip\noindent
  \fbox{\parbox{\dimexpr\linewidth-2\fboxsep-2\fboxrule\relax}{%
  \small\textbf{Answer to #1:} #2}}\par\medskip}

\let\table\tableorg
\let\endtable\endtableorg

\begin{document}

\title[Why Does Structured Output Fail?]{Why Does Structured Output Fail?
An Empirical Study of Prompt-Enforced Pydantic Schemas Across Nine LLMs}

\ifanon
  \author{\fnm{Anonymous} \sur{Author}}
  \affil{\orgname{Anonymous}}
\else
  \author*[1]{\fnm{Amit} \sur{Verma}}\email{amitfromus@gmail.com}
  \affil*[1]{\orgname{Independent Researcher},
    \orgaddress{\city{Phoenix}, \state{AZ}, \country{USA}}}
\fi

\abstract{Pipelines that require structured outputs from hosted language
models often specify a Pydantic schema in the prompt, validate the
response, and retry when validation fails. Although native
structured-output modes provide an alternative, application-side
validation remains important for enforcing custom constraints. Yet
pipeline reliability is often summarized by a single validity rate, which
obscures why outputs fail and which interventions may help. We introduce
a fine-grained taxonomy of schema-conformance failures and a deterministic
classifier that assigns failure categories from model responses and
Pydantic validation errors. We apply this framework to \pn{reported.attempts} generation
attempts across nine OpenAI and Anthropic model/API configurations, using questions from
HotpotQA and FinanceBench. Our results show that similar validity rates
can mask different failure patterns, and that dominant failure categories
vary with schema constraints and task. Markdown-fenced JSON is the most
common failure:    among \pn{fence.reported.n} fenced failures, only \pn{fence.reported.pure_share} pass validation after fence removal, while the remainder conceal additional violations,
predominantly field-length violations. In an exploratory prompt
ablation, variants that suppress fences appear to shift failures toward
length violations, but the shift
is consistent with exposing violations that were already present,
illustrating how aggregate validity rates can obscure changes in failure
composition. We derive four engineering practices for validator, schema,
retry and evaluation design, three supported by our data and one, on retry
feedback, proposed but not tested, and release the classifier,
completions, and annotations.}


\maketitle

\section{Introduction}
\label{sec:intro}

A practitioner building an entity-extraction pipeline needs canonical
identifiers and writes an unremarkable constraint: \idpattern, a
3--31-character lowercase identifier beginning with a letter and
containing letters, digits, or underscores. Asked for the identifier of
an entity whose name is a street address, a GPT-5.6 model in our study
returns \code{888\_7th\_avenue}. The rendering is faithful to the name and
fails the anchor \code{\textasciicircum{}[a-z]}. A conforming identifier exists
(\code{building\_888\_7th\_avenue}), but the task does not specify which
conforming normalisation to use.

Under the evaluation practice common for structured output, this failure
is indistinguishable from a model that emitted no JSON, a model that
returned the schema definition instead of an instance, and a model that
produced a conformant object inside a markdown fence. All four increment
the same counter, although they have different causes and call for
different remedies.

Constrained decoding and vendor structured-output modes prevent many
syntactic failures, but they do not remove the need to validate and
repair. Support for JSON Schema features varies between implementations
\citep{geng2025jsonschemabench}, and Pydantic model validators, such as
cross-field agreement and membership checks, have no JSON Schema
counterpart and cannot be enforced at decoding time. A pipeline that
must run across vendors, model generations, or endpoints therefore
keeps the schema in the prompt and validates afterwards, and its
validate-and-retry loop stays in place even where decoding is
constrained. How that loop fails is an engineering question, and a single validity
rate cannot answer it.

We treat failure diagnosis as part of pipeline design. A validation log
records whether an output passed and the errors Pydantic raised for it.
That record does not separate a formatting failure from a schema
violation it may hide, does not say whether a pattern violation came from
the character class or from a length ceiling written into the pattern,
and does not say whether a change to the prompt or schema removed a
failure or exposed one that was already present. Each distinction changes
what an engineer should fix: the parser, the schema, the retry feedback,
or the way the pipeline is evaluated.

We study the prompt-and-validate paradigm: the schema is stated in the
prompt, the completion is validated with Pydantic, and failures are
retried with the validation error appended. We do not evaluate
constrained decoding (\cref{sec:related}). We ask three questions.
\begin{description}[leftmargin=1.4em,itemsep=1pt,topsep=2pt,font=\normalfont\bfseries]
  \item[RQ1] Do configurations with similar failure rates fail in the
  same way?
  \item[RQ2] Does a failure profile stay stable when the schema's
  constraints or the task domain change?
  \item[RQ3] What do formatting failures conceal, and how does that
  change the reading of repairs and prompt ablations?
\end{description}
We define a taxonomy of conformance failures, assign each failing attempt
a category with a deterministic classifier, and apply it to
\pn{reported.attempts} attempts across \pn{reported.cells} configurations
of nine models from two vendors, one baseline schema and
two targeted variants, and two task domains. A configuration is a model
endpoint called through its vendor's API with the request parameters in
\cref{app:manifest}, so comparisons are between these configurations, not
between model capabilities in general.

The answer to all three questions is that the validity rate hides what an
engineer needs. Configurations with overlapping failure rates fall into
four profiles defined by their dominant category (RQ1). Raising length
caps changes which category dominates for the same model, and revalidating
the same stored outputs against the relaxed schema shows that the caps
alone account for the removed length violations; the same model also
shows different profiles on the two domains (RQ2). Only
\pn{fence.reported.pure_share} of fenced failures validate once the fence
is removed. The rest conceal a length or pattern violation that the
returned parse error never mentions, and ignoring this makes a prompt
ablation appear to move failures from one mode to another (RQ3).

\paragraph{Contributions.}
The central contribution is a failure-diagnosis framework for
prompt-and-validate pipelines: it separates visible output-format
failures from the schema violations beneath them, and uses targeted
interventions to tell failure causes apart. Concretely:
\begin{enumerate}[leftmargin=1.4em,itemsep=1pt,topsep=2pt]
  \item A reproducible diagnostic taxonomy and deterministic classifier,
  with an explicit precedence rule, a development check and a
  pre-registered, blind held-out check against human labels, and released
  code that relabels the corpus from stored completions
  (\cref{sec:setup:taxonomy}).
  \item Evidence that visible formatting failures conceal underlying
  violations: the parse error returned for a fenced output never names
  the violation beneath it, and ignoring concealment misreads both repair
  loops and prompt ablations (\cref{sec:results:concealment}).
  \item Targeted schema interventions that separate failure causes: an
  offline revalidation that holds model output fixed attributes the
  removed length violations to the relaxed caps, and in a small,
  unmatched comparison, loosening an identifier's character class was
  followed by no leading-digit violations while the length overflow
  encoded in the same pattern remained
  (\cref{sec:results:surface}).
  \item Engineering guidance for validator, schema, retry and evaluation
  design, each practice tied to the finding that supports it
  (\cref{sec:implications}).
\end{enumerate}
The results across the nine model configurations
(\cref{sec:results:rates,sec:results:profiles}) are the evidence base for
these contributions, not a ranking of models: similar failure rates hide
different failure profiles, which is why a single rate cannot guide the
design decisions above.

\paragraph{Artifact.} The classifier, harness, locked results database,
raw completions, annotations, and the scripts that regenerate every table
and figure are available at \url{https://github.com/amitvermaknw/schema-enforcement};
the version described in this paper (release \code{emse-submission}) is
archived at \url{https://doi.org/10.5281/zenodo.23586834}
(see also the Statements and Declarations).

\section{Related Work}
\label{sec:related}

\paragraph{Constrained decoding and the scope of this study.}
The main technical alternative to prompt enforcement is constrained
decoding, which masks tokens so that only grammar-conformant
continuations remain. \citet{willard2023outlines} compile regular
expressions into finite-state machines and \citet{dong2024xgrammar} use
pushdown automata for context-free grammars; \citet{geng2025jsonschemabench}
evaluate six frameworks over ten thousand real-world JSON Schemas and find
that support for schema features varies substantially between
implementations. Our exclusion of this paradigm is a statement about what
is deployed, not about what is enforceable: character classes, string
lengths and enumerations are expressible in these formalisms, and we make
no claim that decoder enforcement could not prevent the failures we
report. Vendor structured-output facilities are excluded for the same
reason. They differ by vendor and version in how, and how strictly, they
enforce a schema, and none is the prompt-only configuration we study.

\paragraph{Structured-output benchmarks.}
\citet{tenckhoff2026llmstructbench} evaluate open-weight models, with
GPT-4o as a closed-weight reference, under five prompting configurations
on email-to-JSON extraction. Four of the five set the inference API's
format parameter, to \code{json} or to a JSON Schema, and so add
API-level enforcement; only their prompt-only condition is comparable
to ours. They recover JSON from surrounding text as preprocessing, which
makes wrapping visible in their pipeline but not a category.
\citet{singh2026sob} also draw on HotpotQA but score value accuracy
behind a binary schema gate, grouping conformance failures into parse
failures and schema violations rather than distinguishing them further.
\citet{ferguson2026extractbench} score extracted values against gold
labels with field-dependent correctness. Their unit is the fidelity of a
value; ours is the conformance failure, and we do not evaluate whether
extracted content is correct.

\paragraph{Format adherence.}
\citet{liu2026paraconsist} show that content-preserving paraphrases can
push a model out of a requested response mode into conversational prose
even at temperature zero, the phenomenon our \cat{prose\_preamble} and
\cat{markdown\_wrap} categories measure under a schema.
\citet{xia2024fofo} and \citet{zhou2023ifeval} measure format-following
separately from content quality, and \citet{tam2024speak} find that
format restrictions degrade reasoning even when parse errors are rare.
Their question is what adherence costs; ours is how adherence fails.

\paragraph{Repair and failure taxonomies.}
\citet{huang2024selfcorrect} find that self-correction without an
external signal does not reliably improve reasoning. A Pydantic
validation error is an external, localised signal naming the field and
constraint, so our repair loop is not intrinsic self-correction.
\citet{cemri2025mast} build an empirically grounded taxonomy of
multi-agent system failures validated by multiple annotators. We work at
the level of a single node attempt, with categories assigned by
deterministic predicates and human labels used only to check the
classifier.

\paragraph{Software-engineering perspective.}
A Pydantic schema applied to a model's reply is a runtime contract in the
sense of design by contract \citep{meyer1992contract}: a postcondition
checked at a component boundary. What changes with a language-model
component is the response to a violation. The component cannot be
corrected, only re-invoked, so the information the contract check
returns becomes the input to the next attempt, and its diagnostic
content matters for repair as well as for reporting. Prompts in such
systems are themselves software artifacts: \citet{liang2025prompts} find
that developers treat prompts as programs but build them less
systematically than code, and the literature review of
\citet{hou2024llm4se} catalogues the use of language models across
software-engineering tasks. Closest to our setting,
\citet{shankar2024spade} synthesize data-quality assertions for language
model pipelines from a developer's prompt history; their question is
which checks to write, ours is how to read the failures a fixed set of
checks reports. Fault taxonomies are an established empirical
instrument; \citet{humbatova2020taxonomy} derive one for real faults in
deep learning systems. Our taxonomy is
narrower and operational, assigned by deterministic predicates to runtime
conformance failures rather than by annotators to reported faults. On
diagnosis, \citet{yuan2012logging} find that developers often do not get
log messages right the first time and later spend effort revising them.
A parse error recorded for a fenced output raises a related problem: the
record is accurate but omits the violation an engineer would need to act
on
(\cref{sec:results:concealment}).

\paragraph{Relation to extract-then-validate practice.}
A common engineering response to these failures is to extract the first
JSON object from the completion, validate it, and report syntax errors
separately from schema errors. As usually reported, that split is coarser than ours in two ways that matter
for the results below. When only the post-extraction result is kept, a fenced
payload that also violates a length cap counts as a schema error and the
extraction status is lost, so wrapper--constraint co-occurrence is not
recorded (\cref{sec:results:concealment}); and when schema errors are
reported as one class, a pattern violation caused by length cannot be told
from one caused by the character class (\cref{sec:results:surface}). A
pipeline that retains both is doing the layered analysis of
\cref{sec:results:concealment}; what we add is evidence that discarding
either changes the conclusions drawn.
Our categories refine the schema-error class and keep the formatting
layer visible beneath it.
The contribution is therefore not a finer error vocabulary on its own. It
is keeping both the visible formatting failure and the schema violation
beneath it, and showing on one corpus that discarding either changes an
engineering conclusion: counting fence removal as a repair would report
the compound remainder as solved (\cref{sec:results:concealment}),
reading primary labels alone makes a prompt ablation appear to move
failures between modes (\cref{sec:results:concealment}), and a single
schema-error class hides that loosening a character class leaves a
length overflow encoded in the same pattern untouched
(\cref{sec:results:surface}).

\section{Study Design}
\label{sec:setup}

\subsection{Pipeline, schemas and enforcement}
\label{sec:setup:pipeline}

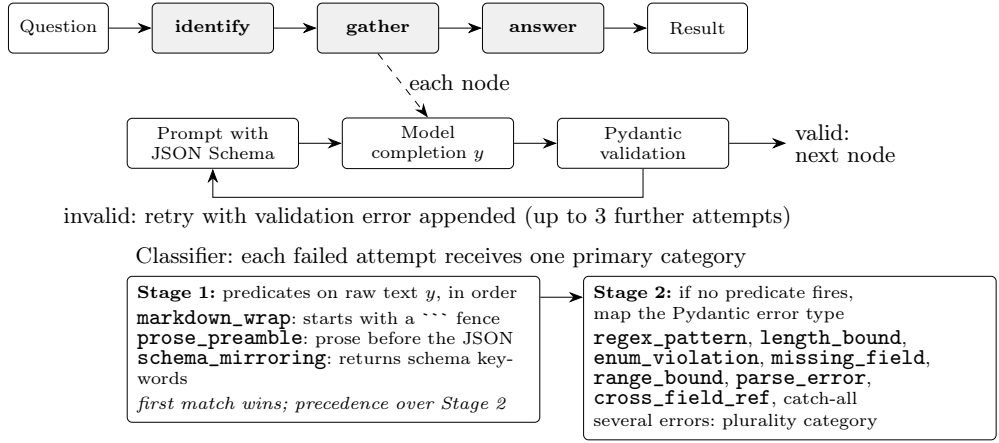
\begin{figure}[!htbp]
\centering
\begin{adjustbox}{max width=\linewidth}
\begin{tikzpicture}[
  font=\footnotesize,
  >={Stealth[length=2mm]},
  node distance=4mm and 6mm,
  box/.style={draw, rounded corners=2pt, align=center, inner sep=3pt,
              minimum height=7mm},
  node/.style={box, fill=black!6, minimum width=17mm},
  stp/.style={box, minimum width=24mm},
  cat/.style={align=left, inner sep=2pt},
  stage/.style={draw, rounded corners=2pt, inner sep=4pt, align=left},
]
\node[box, minimum width=14mm] (q) {Question};
\node[node, right=of q] (n1) {\textbf{identify}};
\node[node, right=of n1] (n2) {\textbf{gather}};
\node[node, right=of n2] (n3) {\textbf{answer}};
\node[box, right=of n3, minimum width=14mm] (out) {Result};
\draw[->] (q) -- (n1);
\draw[->] (n1) -- (n2);
\draw[->] (n2) -- (n3);
\draw[->] (n3) -- (out);

\node[stp, below=9mm of n1] (prompt) {Prompt with\\JSON Schema};
\node[stp, right=of prompt] (llm) {Model\\completion $y$};
\node[stp, right=of llm] (val) {Pydantic\\validation};
\draw[->] (prompt) -- (llm);
\draw[->] (llm) -- (val);
\draw[->, dashed] (n2.south) -- node[right, pos=0.45, font=\scriptsize]{each node} (llm.north);
\draw[->] (val.east) -- ++(8mm,0) node[right, font=\scriptsize, align=left]{valid:\\next node};
\draw[->] (val.south) -- ++(0,-4mm) -| node[pos=0.25, below, font=\scriptsize]
  {invalid: retry with validation error appended (up to 3 further attempts)} (prompt.south);

\node[stage, below=15mm of prompt.south west, anchor=north west, text width=55mm] (s1)
  {\raggedright\textbf{Stage 1:} predicates on raw text $y$, in order\\[2pt]
   \cat{markdown\_wrap}: starts with a \texttt{\textasciigrave\textasciigrave\textasciigrave} fence\\
   \cat{prose\_preamble}: prose before the JSON\\
   \cat{schema\_mirroring}: returns schema keywords\\[2pt]
   \emph{first match wins; precedence over Stage 2}};
\node[stage, right=6mm of s1.north east, anchor=north west, text width=55mm] (s2)
  {\raggedright\textbf{Stage 2:} if no predicate fires,\\map the Pydantic error type\\[2pt]
   \cat{regex\_pattern}, \cat{length\_bound}, \cat{enum\_violation},
   \cat{missing\_field}, \cat{range\_bound}, \cat{parse\_error},
   \cat{cross\_field\_ref}, catch-all\\[1pt]
   several errors: plurality category};
\draw[->] (s1.east |- s2.north west) ++(0,-3mm) -- ++(6mm,0);
\node[font=\scriptsize, anchor=south west] at (s1.north west) {Classifier: each failed attempt receives one primary category};
\end{tikzpicture}
\end{adjustbox}
\caption{Study design. Each question passes through three nodes, each with
its own Pydantic schema stated in the prompt; a failed validation is
retried with the error appended. Every failed attempt is assigned one
primary category, first by behavioural predicates on the raw completion
and otherwise by the Pydantic error type (\cref{sec:setup:taxonomy}).}
\label{fig:overview}
\end{figure}

\Cref{fig:overview} summarises the design. Each run processes one
question through a three-node sequential pipeline.
\textbf{identify} extracts entities with canonical identifiers and types
and designates a primary entity; \textbf{gather} produces facts keyed to
those identifiers with a coverage summary; \textbf{answer} returns the
answer, a numeric confidence, its categorical bucket and supporting
identifiers. Each node has its own Pydantic model. The system prompt
states the schema and asks for JSON only; the completion is validated
with \code{model\_validate\_json} in Pydantic's default lax mode (type
coercion allowed, extra fields ignored; no model sets
\code{model\_config}); on failure the node is retried with the
validation error appended, up to three further times. An \emph{attempt} is
one generation against one node's schema, including retries, so attempts
per run vary with failure rate: on HotpotQA, from
\pn{c.claude-opus-4-8@medium@hotpotqa.attempts_per_run} for Opus~4.8 to
\pn{c.claude-haiku-4-5@medium@hotpotqa.attempts_per_run} for Haiku~4.5.

The \emph{medium} schema tier carries the constraints analysed
throughout: the identifier pattern \idpattern{} on entity and fact
identifiers, enumerations on entity and question type, character caps
on free text (\code{coverage\_notes} 250, \code{answer} 400, \code{fact}
400), list-length bounds and numeric ranges. Model-level validators add
four cross-field checks---primary-entity membership and entity-count
agreement in identify, fact-count agreement in gather, and
confidence-bucket coherence in answer---and, in answer, a per-item
identifier pattern check on \code{supporting\_entity\_ids}. The system
prompt carries each node's JSON Schema, so field-level constraints and
field descriptions are visible from the first attempt. The four
cross-field checks are stated in field descriptions; the
supporting-identifier check is stated nowhere in the prompt and becomes
visible only through retry feedback (\cref{tab:visibility}).
Two targeted variants of the baseline schema each change one constraint
family. \emph{medium\_v2} raises the caps to 1000, 1200 and 800 and is
run on Sonnet~4.6 only; it also omits the three node-level docstrings,
which the prompt carries as schema descriptions but which state no
constraint. \emph{medium\_v3\_loose\_regex} loosens the identifier
pattern to \idpatternloose{}, keeping the 31-character ceiling, rewords
the \code{canonical\_id} description to match, and is run on
gpt-5.6-sol and gpt-5.6-terra only. The design is therefore not fully
crossed; \cref{app:manifest} lists every cell.

\subsection{Models, data and reporting basis}
\label{sec:setup:basis}

Nine models from two vendors: \code{gpt-4o}, \code{gpt-4o-mini},
\code{gpt-5.6-luna}, \code{gpt-5.6-sol}, \code{gpt-5.6-terra},
\code{claude-haiku-4-5-20251001}, \code{claude-sonnet-4-6},
\code{claude-sonnet-5} and \code{claude-opus-4-8}, all called through the
vendors' own APIs between \pn{collection.date_min} and
\pn{collection.date_max}. Only the Haiku request names a dated snapshot;
the two gpt-4o models resolved to dated snapshots, and the remaining
endpoints returned the requested alias, which identifies the endpoint but
not an immutable model version (\cref{app:manifest}). HotpotQA
\citep{yang2018hotpotqa} questions are the first records of the
validation split (distractor configuration) and FinanceBench
\citep{islam2023financebench} questions the first records of the public
sample released as \code{PatronusAI/financebench}, which is a subset of the
full FinanceBench question set and has a single \code{train} split, with
evidence truncated to 2{,}000 characters; neither is shuffled,
so results describe these fixed question sets. Most cells contain about
one run per question; the Haiku cells repeat questions, and its
FinanceBench cell covers
\pn{c.claude-haiku-4-5@medium@financebench.questions} distinct questions
(\cref{app:manifest}). Claude Sonnet~4.6 and gpt-4o were run on HotpotQA
only. We request \code{temperature=0} where supported; the GPT-5.6 family,
Sonnet~5 and Opus~4.8 reject the parameter and run at provider defaults.
The Anthropic branch sets \code{max\_tokens} to 4{,}096 and the OpenAI
branch sets none. Resolved model versions and sampling details are in
\cref{app:manifest}.

A \emph{cell}, or configuration, is one combination of model, schema
tier and task domain under the baseline prompt. \Cref{tab:manifest} has
one row per cell plus one row for each of the two prompt variants,
which share the Haiku HotpotQA cell and add no cells.

Two conventions define the reporting basis. Cells with fewer than 30
baseline runs are excluded as pilots, a threshold fixed before analysis,
and the \pn{excluded_variant.runs} prompt-variant runs, which share the
Haiku HotpotQA configuration, are excluded everywhere except
\cref{sec:results:concealment}. The basis contains
\pn{reported.attempts} attempts and \pn{reported.failures} failures in
\pn{reported.cells} configurations, from a full corpus of
\pn{full.attempts} attempts in \pn{full.cells} cells. Classifier
validation (\cref{sec:setup:taxonomy}) uses the full corpus and says so.

\paragraph{Uncertainty.}
Attempts are not independent: retries follow failures, questions repeat
in some cells, and later nodes are reached only after earlier ones
succeed. Intervals are therefore from a question-clustered bootstrap
(\pn{meta.bootstrap_reps} replicates, percentile 95\%) that resamples
questions within each dataset and carries every run and attempt of a
selected question. They describe variability over the observed question
set and do not make the fixed prefixes representative of the benchmarks.
Because a failure adds an attempt under retry-until-valid, failures and
attempts co-vary within a question, so for high-failure models the
clustered interval on the failure rate can be narrower than a binomial
interval over attempts. Where every replicate gives the same value, as for
a share of 0 or 1, we report no interval.

\subsection{Taxonomy and classifier}
\label{sec:setup:taxonomy}

The classifier runs only on attempts that failed validation and assigns
one primary category in two stages. \emph{Stage~1} applies behavioural
predicates to the raw completion $y$, in order: \cat{markdown\_wrap} if
$y$ begins, after leading whitespace, with a triple-backtick fence;
\cat{prose\_preamble} if $y$ begins with neither a brace, a bracket nor a
fence, contains a brace-delimited span, and has at least 20 characters of
prose before it; and \cat{schema\_mirroring} if $y$ parses to an object
whose root keys include at least two JSON~Schema meta-keywords such as
\code{properties} and \code{required}. \emph{Stage~2}, reached only if no
predicate fires, maps Pydantic error types: pattern mismatches to
\cat{regex\_pattern}, length violations to \cat{length\_bound}, literal
errors to \cat{enum\_violation}, missing fields to \cat{missing\_field},
numeric bounds to \cat{range\_bound} and \code{json\_invalid} to
\cat{parse\_error}. The map sends \code{int\_type} and \code{float\_type} to
\cat{range\_bound} and the string, list, dict and boolean type errors to
\cat{missing\_field}; any unmapped type, including the lax-mode parsing
errors such as \code{float\_parsing}, falls to the catch-all
\cat{other\_validation\_error}. These type assignments are a coarse
fallback in the released classifier, not a claim that a type error is a
range or missing-field violation. No type or parsing error occurs in the
corpus, in stored error details or beneath stripped fences, so no
reported label depends on this part of the map and it is untested here;
a dedicated \cat{type\_mismatch} category is left to future work. Validator errors arrive as \code{value\_error}; the
supporting-identifier check routes to \cat{regex\_pattern} and cross-field
messages to \cat{cross\_field\_ref}. With several errors, the primary
category is the plurality, ties broken by first occurrence.
Labels follow how a constraint is encoded, not what it intends: an
identifier longer than 31 characters violates the quantifier in
\idpattern{} and is labelled \cat{regex\_pattern}, whereas the same
limit expressed as \code{max\_length} would be labelled
\cat{length\_bound}.

We give Stage-1 predicates precedence to preserve behaviours that are
invisible to Pydantic, which sees only their downstream symptoms. The
precedence has a consequence we exploit in
\cref{sec:results:concealment}: a fenced output that also violates a
constraint is labelled \cat{markdown\_wrap}, so that category measures
strict-parsing failure, not the absence of a further defect.
\Cref{tab:categories} gives the observed distribution;
\pn{reported.categories_observed} of the classifier's eleven categories
occur, and \cat{cross\_field\_ref}, \cat{missing\_field},
\cat{range\_bound} and the catch-all category do not.

\begin{table}[!htbp]
\centering\small
\caption{Primary categories on the reporting basis.}
\label{tab:categories}
\begin{tabular}{@{}lrr@{}}
\toprule
Category & $n$ & Share \\
\midrule
\cat{markdown\_wrap} & 681 & 60.5\% \\
\cat{length\_bound} & 263 & 23.4\% \\
\cat{regex\_pattern} & 126 & 11.2\% \\
\cat{schema\_mirroring} & 38 & 3.4\% \\
\cat{prose\_preamble} & 9 & 0.8\% \\
\cat{enum\_violation} & 8 & 0.7\% \\
\cat{parse\_error} & 1 & 0.1\% \\
\midrule
Total & 1{,}126 &  \\
\bottomrule
\end{tabular}

\end{table}

\paragraph{Validation.}
We checked the classifier against human labels twice. During development
we hand-labelled \pn{validation.n} failing attempts stratified by
predicted category, drawn from the full corpus of \pn{full.failures}
failures so that rare categories had support. The labelling screen
displayed the classifier's label, and the author consulted an LLM
assistant when interpreting outputs; every label was assigned by the
author. This labelling surfaced a detection-order defect in an earlier
classifier, in which Pydantic mapping preceded the behavioural
predicates; we corrected the order and reclassified the corpus. Against
the corrected classifier, unadjudicated agreement is
\pn{validation.blind_agree} of \pn{validation.n}
(\pn{validation.blind_rate}, Wilson 95\% CI \pn{validation.blind_ci}),
and on adjudication all four disagreements were labelling errors. Because
the classifier's label was visible and these samples shaped the
classifier, this is a development-time consistency check, not
independent validation.

After freezing the classifier we ran a pre-registered held-out check:
\pn{heldout.n} further failures from the reported basis, excluding every
development item and stratified over the three categories with attempts
remaining, labelled by hand by the author with the classifier's label
hidden and without AI assistance on any item. Agreement is
\pn{heldout.agree} of \pn{heldout.n} (\pn{heldout.rate}, Wilson 95\% CI
\pn{heldout.ci}; \pn{heldout.weighted_rate} weighted by category
frequency). Checked post hoc against the validator's error record and the
raw completion, none of the \pn{heldout.disagree} disagreements showed
the classifier misapplying the category definitions or the precedence
rule. In \pn{heldout.dis.length_unverified} the annotator could not
confirm from the displayed output a length overflow the validator
reported; \pn{heldout.dis.fence_underlying} fenced outputs were labelled
by the violation beneath the fence and \pn{heldout.dis.fence_ambiguous}
more were marked ambiguous; and \pn{heldout.dis.identifier_pattern}
identifier pattern mismatches were labelled otherwise,
\pn{heldout.dis.identifier_as_enum} of them as \cat{enum\_violation}, the
confusion behind three of the four development disagreements. Stage-2
labels are a fixed map from validator errors, so human labels chiefly
test the category definitions and the Stage-1 predicates, and unaided
reading cannot verify a length label. The gap between the two figures is
consistent with anchoring on the displayed label in the development
round. \Cref{app:classifier} gives per-category results and corpus
screens for the Stage-1 categories.
\section{Results}
\label{sec:results}

\subsection{Failure rates and run completion}
\label{sec:results:rates}

\Cref{tab:models} gives per-model results. The attempt-weighted failure
rate over the reporting basis is \pn{pooled.reported.rate} (95\% CI
\pn{pooled.reported.ci}), but Haiku~4.5 contributes
\pn{g.claude-haiku-4-5@medium.failures} of the \pn{reported.failures}
failures. Excluding it gives \pn{reported.exhaiku.rate}, and the
unweighted mean of the nine medium-tier rates is
\pn{reported.medium.macro_rate}. Attempt-weighted aggregates over this
corpus are therefore largely statements about one model, and we report
per-model figures. No conclusion below depends on ordering models whose
intervals overlap.

\begin{table}[!htbp]
\centering\small
\caption{Per-model results at the medium tier, baseline prompt, task
domains pooled (gpt-4o, gpt-4o-mini and Sonnet~4.6 are HotpotQA only on
the reporting basis). Intervals are question-clustered bootstrap 95\%
intervals (\cref{sec:setup:basis}). Rows are ordered by failure rate for
readability; configurations with overlapping intervals are not ranked.}
\label{tab:models}
\setlength{\tabcolsep}{5pt}
\adjustbox{max width=\textwidth}{\begin{tabular}{lrrrrlllr}
\toprule
Model & Runs & Att. & Fail. & Rate \% & 95\% CI & Dominant & Share \% & 95\% CI \\
\midrule
\code{claude-haiku-4-5} & 166 & 1329 & 878 & 66.1 & [64.9, 67.3] & markdown\_wrap & 72.9 & [70.4, 75.3] \\
\code{claude-sonnet-5} & 82 & 313 & 67 & 21.4 & [17.2, 25.2] & markdown\_wrap & 61.2 & [49.3, 71.4] \\
\code{gpt-4o} & 70 & 250 & 40 & 16.0 & [11.9, 20.0] & schema\_mirroring & 72.5 & [58.8, 87.2] \\
\code{claude-sonnet-4-6} & 50 & 178 & 28 & 15.7 & [10.2, 21.5] & length\_bound & 64.3 & [48.1, 80.8] \\
\code{gpt-4o-mini} & 70 & 242 & 33 & 13.6 & [8.0, 19.7] & regex\_pattern & 66.7 & [45.7, 83.3] \\
\code{gpt-5.6-terra} & 84 & 266 & 14 & 5.3 & [3.1, 7.7] & regex\_pattern & 85.7 & [64.3, 100.0] \\
\code{claude-opus-4-8} & 84 & 265 & 13 & 4.9 & [2.7, 7.4] & length\_bound & 76.9 & [53.8, 100.0] \\
\code{gpt-5.6-luna} & 85 & 266 & 11 & 4.1 & [2.2, 6.3] & regex\_pattern & 72.7 & [44.4, 100.0] \\
\code{gpt-5.6-sol} & 82 & 255 & 9 & 3.5 & [1.6, 5.7] & regex\_pattern & 88.9 & [62.5, 100.0] \\
\bottomrule
\end{tabular}
}
\end{table}

Attempt failure and pipeline completion are different quantities. Under
the repair loop, \pn{pooled.reported.completed} of
\pn{pooled.reported.runs} runs complete all three nodes. Every run
completes for seven of the nine models; gpt-4o-mini completes
\pn{g.gpt-4o-mini@medium.completed} of \pn{g.gpt-4o-mini@medium.runs}
(its only incomplete run exhausted its retries on the supporting-identifier
check, the one enforced constraint not stated in the prompt;
\cref{app:crossfield}),
and Haiku~4.5, whose attempt failure rate is
\pn{g.claude-haiku-4-5@medium.rate}, completes
\pn{g.claude-haiku-4-5@medium.completed} of
\pn{g.claude-haiku-4-5@medium.runs} at
a cost of
\pn{g.claude-haiku-4-5@medium.attempts_per_completed_run} attempts per
completed run, counting the attempts of its incomplete runs as well
(\cref{app:breakdowns}).

First attempts, before any feedback, give a stricter view. Of the
\pn{pooled.reported.att0.n} first attempts on the reporting basis,
\pn{pooled.reported.att0.valid} validated
(\pn{pooled.reported.fa.valid_rate}); counting pure-formatting failures
as recovered raises this to \pn{pooled.reported.fa.valid_after_fence_rate}.
None of Haiku~4.5's \pn{g.claude-haiku-4-5@medium.fa.reached} first
attempts validated; per-model first-attempt breakdowns are in
\cref{app:breakdowns}.

\subsection{Dominant-category profiles}
\label{sec:results:profiles}

\begin{figure}[!htbp]
\centering
\includegraphics[width=0.92\linewidth]{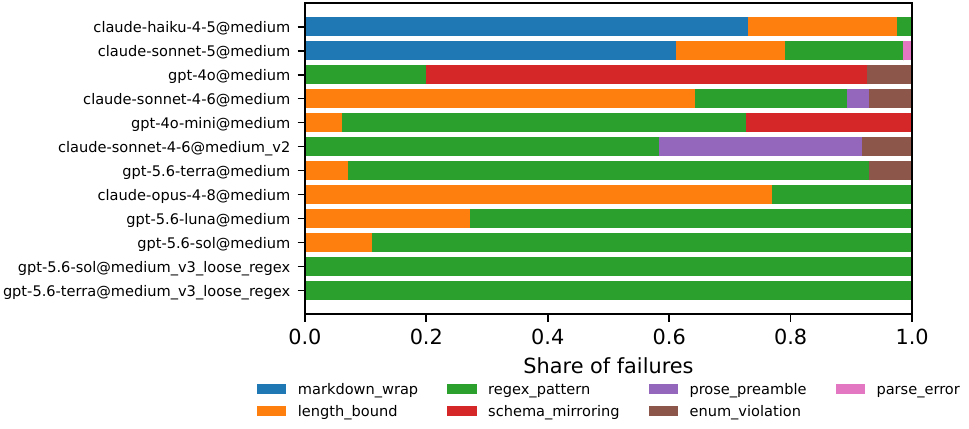}
\caption{Composition of failures by model and schema tier, reporting
basis, ordered by failure rate. Per-model failure counts and rates are
in \cref{tab:models}; counts for every cell, including the schema
variants, are in \cref{tab:manifest}.}
\label{fig:profiles}
\end{figure}

Grouping models by dominant category gives four observed profiles
(\cref{fig:profiles}): \cat{markdown\_wrap} for Haiku~4.5 and Sonnet~5;
\cat{regex\_pattern} for the three GPT-5.6 models and gpt-4o-mini;
\cat{length\_bound} for Opus~4.8 and Sonnet~4.6; and
\cat{schema\_mirroring} for gpt-4o. These are profiles read off the
dominant category, not clusters from any statistical procedure. They are
not determined solely by vendor, model tier, or failure rate: gpt-4o and
gpt-4o-mini share a generation but not a profile, and gpt-4o-mini,
Sonnet~4.6 and gpt-4o fail at \pn{g.gpt-4o-mini@medium.rate},
\pn{g.claude-sonnet-4-6@medium.rate} and \pn{g.gpt-4o@medium.rate}, with
overlapping intervals, under three different dominant categories.
Dominant shares for low-rate models rest on few failures; Opus~4.8's
\pn{g.claude-opus-4-8@medium.dominant_share} share, for example, rests on
\pn{g.claude-opus-4-8@medium.failures} failures (95\% CI
\pn{g.claude-opus-4-8@medium.share.length_bound.ci}).

Some categories are confined to particular models in our data.
\cat{schema\_mirroring} accounts for \pn{g.gpt-4o@medium.dominant_share}
of gpt-4o failures and was not observed in any GPT-5.6 configuration we
tested. \cat{markdown\_wrap} was not observed for Sonnet~4.6 across
\pn{m.claude-sonnet-4-6.attempts} attempts or Opus~4.8 across
\pn{m.claude-opus-4-8.attempts}, yet it accounts for
\pn{g.claude-sonnet-5@medium.dominant_share} of Sonnet~5 failures and
\pn{g.claude-haiku-4-5@medium.dominant_share} of Haiku~4.5 failures.
Haiku~4.5 and gpt-4o are each dominant in a category the other did not
produce in our observations. Among the
\pn{divergence.groups_qualifying} model--tier groups with at least 20
failures, the largest difference between two groups in the share of a
single category is \pn{divergence.top.0.pp} percentage points, attained
by \pn{divergence.top.n_tied} pairs, each setting Haiku~4.5 against a
group with no \cat{markdown\_wrap} failures (\cref{app:divergence}).

Profiles also vary with the task distribution. Sonnet~5 fails at nearly
the same rate on both domains (\pn{c.claude-sonnet-5@medium@financebench.rate}
on FinanceBench, \pn{c.claude-sonnet-5@medium@hotpotqa.rate} on HotpotQA),
but on HotpotQA \cat{markdown\_wrap} accounts for
\pn{c.claude-sonnet-5@medium@hotpotqa.dominant_share} of its
\pn{c.claude-sonnet-5@medium@hotpotqa.failures} failures, while on
FinanceBench no category clearly dominates its
\pn{c.claude-sonnet-5@medium@financebench.failures}:
\cat{length\_bound} \pn{c.claude-sonnet-5@medium@financebench.share.length_bound}
(95\% CI \pn{c.claude-sonnet-5@medium@financebench.share.length_bound.ci}),
\cat{regex\_pattern} \pn{c.claude-sonnet-5@medium@financebench.share.regex_pattern}
and \cat{markdown\_wrap} \pn{c.claude-sonnet-5@medium@financebench.share.markdown_wrap}.
The observed profile is thus a property of model, constraints, prompt and
task distribution together. Opus~4.8 also changes dominant category
across domains, but its HotpotQA profile rests on
\pn{c.claude-opus-4-8@medium@hotpotqa.failures} failures and we do not
interpret it (\cref{app:breakdowns}).

\rqanswer{RQ1}{No. Configurations with overlapping failure rates fall
into four profiles with different dominant categories, and some categories
occur only in particular configurations. A validity rate therefore does
not identify which failure an engineer must fix.}

\subsection{Profiles depend on the constraint surface}
\label{sec:results:surface}

\begin{table}[!htbp]
\centering\small
\caption{Sonnet~4.6, HotpotQA. Top: composition at the medium and
medium\_v2 tiers. Bottom: the stored medium outputs revalidated against
the medium\_v2 schema, beside the fresh medium\_v2 generations. The
fresh cell differs in size and retry structure and is not paired by
question.}
\label{tab:calibration}
\begin{tabular}{@{}lrrrr@{}}
\toprule
& \multicolumn{2}{c}{medium (178 att.)} & \multicolumn{2}{c}{medium\_v2 (324 att.)} \\
\cmidrule(lr){2-3}\cmidrule(lr){4-5}
Category & $n$ & rate & $n$ & rate \\
\midrule
\cat{length\_bound} & 18 & 10.1\% & 0 & 0.0\% \\
\cat{regex\_pattern} & 7 & 3.9\% & 14 & 4.3\% \\
\cat{prose\_preamble} & 1 & 0.6\% & 8 & 2.5\% \\
\cat{enum\_violation} & 2 & 1.1\% & 2 & 0.6\% \\
\midrule
All failures & 28 & 15.7\% & 24 & 7.4\% \\
\bottomrule
\end{tabular}

\vspace{4pt}
\begin{tabular}{lrrr}
\toprule
Analysis & Attempts & Failures & Rate \% \\
\midrule
Original medium outputs, medium schema & 178 & 28 & 15.7 \\
Same outputs, medium\_v2 schema & 178 & 10 & 5.6 \\
Fresh medium\_v2 generations & 324 & 24 & 7.4 \\
\bottomrule
\end{tabular}

\end{table}

A profile is not a property of the model alone. Sonnet~4.6 is
\cat{length\_bound}-dominant at the medium tier
(\pn{calib.medium.length_bound} of \pn{calib.medium.failures} failures).
With the caps raised in medium\_v2, \cat{length\_bound} falls to
\pn{calib.medium_v2.length_bound}, \cat{regex\_pattern} becomes dominant
(\pn{calib.medium_v2.regex_pattern} of \pn{calib.medium_v2.failures}), and
the failure rate moves from \pn{calib.medium.rate} to
\pn{calib.medium_v2.rate} (\cref{tab:calibration}).

The error locations identify which caps were binding:
\pn{calib.field.coverage_notes.n} \code{coverage\_notes} values of
\pn{calib.field.coverage_notes.min}--\pn{calib.field.coverage_notes.max}
characters against the 250 cap (mean overrun
\pn{calib.field.coverage_notes.mean_overrun}, maximum
\pn{calib.field.coverage_notes.max_overrun}), \pn{calib.field.answer.n}
\code{answer} values of
\pn{calib.field.answer.min}--\pn{calib.field.answer.max} against 400 (mean
overrun \pn{calib.field.answer.mean_overrun}), and
\pn{calib.field.fact.n} \code{fact} value of
\pn{calib.field.fact.max} characters against 400. These caps sat below
the lengths the model produced, not far below them. Because medium\_v2 is
a fresh generation, its result could in principle reflect changed output
rather than changed caps. We therefore revalidated all \pn{cf.n} stored
medium outputs, valid and failed, against the medium\_v2 schema. All
\pn{cf.invalid_to_valid.length_bound} \cat{length\_bound} failures pass,
no valid output becomes invalid, and the remaining
\pn{cf.samev2.failures} failures are the original \cat{regex\_pattern},
\cat{enum\_violation} and \cat{prose\_preamble} cases. Relaxing the caps
is sufficient to eliminate the observed length violations. This is an
offline revalidation, not a simulated medium\_v2 run: under the relaxed
schema the retries that followed those failures would not have occurred.

Loosening the identifier character class is the contrasting case. On
HotpotQA, gpt-5.6-sol and gpt-5.6-terra produced
\pn{pert.medium.identifiers} violating identifiers under the baseline
pattern and \pn{pert.medium_v3_loose_regex.identifiers} under the
loosened pattern. Leading-digit violations such as the example in
\cref{sec:intro} went from \pn{pert.medium.leading_digit} to
\pn{pert.medium_v3_loose_regex.leading_digit}, as the loosened class
permits them. Length overflows against the retained 31-character ceiling
occurred under both patterns (\pn{pert.medium.length_overflow} and
\pn{pert.medium_v3_loose_regex.length_overflow}); with a handful of
failures per cell we do not interpret that change or the cell rates
(\cref{app:perturbation}). No leading-digit violations were observed
under the loosened pattern, while length overflow, which the loosened
class does not address, still occurred. Because these runs are fresh
generations rather than a revalidation of fixed outputs, and the counts
are small, we read this as consistent with a sub-cause-specific effect,
not as an estimate of one. The \cat{regex\_pattern} label conceals the
overflow because the ceiling is written into the pattern's quantifier.

\rqanswer{RQ2}{No. Raising length caps changes the dominant category for
the same model, and revalidating the same stored outputs attributes the
removed length violations to the caps alone. Under a loosened identifier
character class, no leading-digit violations were observed
(\pn{pert.medium.leading_digit} to
\pn{pert.medium_v3_loose_regex.leading_digit}) while length overflows
remained; with counts this small and unmatched runs, this is an
observation, not a measured effect. The same model
also shows different profiles on the two task domains
(\cref{sec:results:profiles}).}

\subsection{Formatting failures conceal underlying schema violations}
\label{sec:results:concealment}

\begin{table}[!htbp]
\centering\small
\caption{Fence-strip revalidation of \cat{markdown\_wrap} failures,
reporting basis. \emph{Pure formatting} validates once the fence is
removed; \emph{compound} raises a further error. Percentages are
rounded independently and may not sum exactly.}
\label{tab:fence}
\begin{tabular}{@{}lrrr@{}}
\toprule
Model & Wrapped & Pure formatting & Compound \\
\midrule
\code{claude-haiku-4-5} & 640 & 290 (45.3\%) & 350 (54.7\%) \\
\code{claude-sonnet-5} & 41 & 37 (90.2\%) & 4 (9.8\%) \\
\midrule
All wrapped & 681 & 327 (48.0\%) & 354 (52.0\%) \\
\bottomrule
\end{tabular}

\end{table}

Because \cat{markdown\_wrap} takes precedence, it says nothing about what
lies inside the fence. We stripped the fence from every wrapped failure
and revalidated the payload (\cref{tab:fence}).
\pn{fence.reported.pure_share} validate; the rest raise a further error,
overwhelmingly on length: \pn{fence.reported.concealed.length_bound}
\cat{length\_bound}, \pn{fence.reported.concealed.regex_pattern}
\cat{regex\_pattern} and \pn{fence.reported.concealed.missing_field}
\cat{missing\_field}. The two wrapping models share a dominant category
but not what the wrapper hides: \pn{g.claude-sonnet-5@medium.fence_compound}
of Sonnet~5's \pn{g.claude-sonnet-5@medium.fence_n} wrapped failures are
compound (95\% CI \pn{g.claude-sonnet-5@medium.fence_compound_ci}),
against \pn{g.claude-haiku-4-5@medium.fence_compound_share} of
Haiku~4.5's \pn{g.claude-haiku-4-5@medium.fence_n} (95\% CI
\pn{g.claude-haiku-4-5@medium.fence_compound_ci}). The contrast is not a
domain-mix artefact: on HotpotQA alone the compound shares are
\pn{c.claude-sonnet-5@medium@hotpotqa.fence_compound_share} and
\pn{c.claude-haiku-4-5@medium@hotpotqa.fence_compound_share}, and Haiku's
rises to \pn{c.claude-haiku-4-5@medium@financebench.fence_compound_share}
on FinanceBench. Counting co-occurrence, \cat{length\_bound} reaches
\pn{reported.length_bound_incl_concealed} instances against a primary
count of \pn{reported.cat.length_bound}, so single-primary-label counts
omit some detected co-occurring violations.

\paragraph{What the repair loop repairs.}
Concealment is visible in what retries left unrepaired. All
\pn{g.claude-haiku-4-5@medium.fa.reached} of Haiku~4.5's first attempts
failed strict validation, every one of them fenced; we audited the
extraction (one first attempt per run-node, contiguous attempt numbers,
no attempts after a valid one, and fences confirmed on the raw text;
\cref{app:breakdowns}). On the second attempt, after the validation error
was returned, \pn{g.claude-haiku-4-5@medium.att1.valid} of
\pn{g.claude-haiku-4-5@medium.att1.n} validated and fences persisted in
\pn{g.claude-haiku-4-5@medium.att1.cat.markdown_wrap}; no later attempt
was fenced. What remained was length: \cat{length\_bound} accounts for
every failure from the third attempt onward, and all
\pn{g.claude-haiku-4-5@medium.exhausted} run-nodes that exhausted their
retries ended on \cat{length\_bound}, \pn{g.claude-haiku-4-5@medium.exhausted.node.gather}
of them in gather. These are exactly the runs that did not complete. Within the loop, nearly all of this model's failures
were repaired (\pn{g.claude-haiku-4-5@medium.fa.retry_success} of
\pn{g.claude-haiku-4-5@medium.fa.failed} failed first attempts eventually
validated), and its residual failures were the violations the fence had concealed. In
\pn{conceal.exh.same_field} of the \pn{conceal.exh.n} exhausted run-nodes,
the fence-stripped first attempt already violated the cap of the same field
as the final attempt, and in \pn{conceal.exh.persist} that field exceeded
its cap on every attempt. The error returned for each of their
\pn{conceal.exh.fenced} fenced attempts was a JSON parse error that did not
mention the cap, so \pn{conceal.exh.len_feedback_once} of these run-nodes
received the length error once before exhausting their retries and
\pn{conceal.exh.len_feedback_twice} received it twice. Offline on the stored attempts, counting pure-formatting
failures as recovered leaves \pn{pooled.reported.residual_rate} of
attempts invalid (95\% CI \pn{pooled.reported.residual_ci}) against
\pn{pooled.reported.rate} strict, \pn{g.claude-haiku-4-5@medium.residual_rate}
for Haiku~4.5 and \pn{g.claude-sonnet-5@medium.residual_rate} for
Sonnet~5. These are not outcomes of rerunning the pipeline with fence
removal enabled, which would change retries and downstream exposure.

\begin{table}[!htbp]
\centering\small
\caption{Exploratory prompt ablation on Haiku~4.5, HotpotQA, rates over
attempts. The last row counts length violations that are primary or
concealed beneath a fence. Design limits are given in
\cref{sec:results:concealment}.}
\label{tab:ablation}
\begin{tabular}{@{}lrrr@{}}
\toprule
& Baseline & Aggressive & Few-shot \\
\midrule
Attempts & 1{,}066 & 189 & 176 \\
\cat{markdown\_wrap} & 49.0\% & 33.9\% & 25.0\% \\
\cat{length\_bound} (primary) & 14.4\% & 16.4\% & 22.7\% \\
\cat{regex\_pattern} & 1.5\% & 2.1\% & 2.3\% \\
Failure rate & 64.9\% & 52.4\% & 50.0\% \\
\midrule
Length violations, primary + concealed & 37.8\% & 30.2\% & 36.9\% \\
\bottomrule
\end{tabular}

\end{table}

Concealment also changes how a prompt ablation reads. We use an
exploratory ablation to show this, not to compare prompts: three
system-prompt variants on Haiku~4.5 on HotpotQA, namely the baseline, an
aggressive format instruction, and a few-shot prompt
(\cref{tab:ablation}). The variant cells are much smaller than the
baseline cell, were run immediately after the final baseline batch rather
than interleaved with it, and are not paired by question, so differences
between conditions may reflect question composition or collection time as
well as prompt wording. The wrapping rate is
\pn{ablation.baseline.markdown_wrap} of attempts under the baseline and
\pn{ablation.few_shot.markdown_wrap} under the few-shot prompt, and the
failure rate \pn{ablation.baseline.failure_rate} and
\pn{ablation.few_shot.failure_rate}. The primary \cat{length\_bound} rate
is \pn{ablation.baseline.length_bound} and
\pn{ablation.few_shot.length_bound}, which read alone suggests failures
migrating from wrapping to length. But
a suppressed fence can expose a length violation that was already
present. Counting primary and concealed length violations together gives
\pn{ablation.baseline.length_bound_incl_concealed},
\pn{ablation.aggressive.length_bound_incl_concealed} and
\pn{ablation.few_shot.length_bound_incl_concealed}: the combined rate does
not rise as wrapping falls, so the rise in the primary rate requires no
increase in over-long output. This is an
accounting argument within the observed attempts, and it does not depend
on the prompts having caused the differences between conditions. We do
not interpret the differences among the combined rates, including the
lower aggressive figure, and we draw no conclusion about which prompt is
more effective.

\rqanswer{RQ3}{Only \pn{fence.reported.pure_share} of fenced failures
validate once the fence is removed; the rest conceal a length or pattern
violation that the returned parse error never mentions. Counting concealed
violations changes the reading of the repair loop, whose residual failures
were the violations the fence had concealed, and of the exploratory prompt
ablation, whose apparent shift towards length violations requires no
increase in over-long output.}
\section{Implications for Practice}
\label{sec:implications}
\label{sec:discussion} 

Each practice below follows from a specific finding and targets one
component of a prompt-and-validate pipeline: the validator, the schema,
the retry loop, or the way the pipeline is evaluated
(\cref{tab:practices}). The first, second and fourth are supported at the
scale of this corpus; the third is a design implication of our findings
that we did not test.

\begin{table}[!htbp]
\centering\small
\caption{Engineering practices, the pipeline component each targets, and
the finding that supports it.}
\label{tab:practices}
\begin{adjustbox}{max width=\textwidth}
\begin{tabular}{@{}p{3.1cm}p{4.6cm}p{4.4cm}l@{}}
\toprule
Component & Finding & Practice & Evidence \\
\midrule
Validator & Only \pn{fence.reported.pure_share} of fenced failures
validate once the fence is removed & Strip fences, revalidate, and record
the residual violation & \cref{sec:results:concealment} \\
Schema & A length ceiling written into a pattern is reported as a
pattern violation; caps below observed lengths drive the dominant
category & Express length limits separately from patterns; measure
output lengths before fixing caps & \cref{sec:results:surface} \\
Retry loop & The parse error for a fenced output never names the
violation beneath it & Build retry feedback from the fence-stripped
validation (untested) & \cref{sec:results:concealment} \\
Evaluation & Aggregate rates are dominated by one configuration;
primary labels misread a prompt ablation & Report failure composition
per configuration, counting concealed violations &
\cref{sec:results:rates,sec:results:concealment} \\
\bottomrule
\end{tabular}
\end{adjustbox}
\end{table}

\paragraph{Validator: strip fences, then revalidate and record what remains.}
Fence removal is a partial repair: it recovers the pure-formatting share
of wrapped failures, which ranged from most of Sonnet~5's to under half
of Haiku~4.5's (\cref{sec:results:concealment}). A pipeline that counts
fence removal as a fix would report the compound remainder as solved.

\paragraph{Schema: separate conflated causes and set caps from data.}
A \code{\{2,30\}} quantifier turns an over-long identifier into a pattern
violation, and the natural remedy for a pattern violation, loosening the
character class, leaves it untouched (\cref{sec:results:surface}). A
separate \code{max\_length} separates the causes at validation time.
Caps should likewise be set from a field's observed length distribution.
Sonnet~4.6's dominant mode came from caps sitting modestly below its
output lengths, and Haiku~4.5's incomplete runs all ended on length. Such
a cap may still be a real downstream requirement; the measurement
identifies the mismatch the application must resolve, not that the cap
is wrong.

\paragraph{Retry loop: give the model the violation it must fix.}
The error returned for a fenced output is a JSON parse error, so the
repair loop learns of a concealed violation only on a later attempt.
Across the reported basis, \pn{retrygap.concealed_msgs} of the
\pn{retrygap.feedback_msgs} retry messages sent
(\pn{retrygap.concealed_share}) named only a parse error while the
fence-stripped payload already violated the schema, almost all of them
Haiku~4.5's (\pn{retrygap.model.claude-haiku-4-5.concealed_msgs}) and
\pn{retrygap.cause.length} of them concealing a length violation; they
span \pn{retrygap.runnodes} run-nodes, including all
\pn{retrygap.runnodes_exhausted} that exhausted their retries
(\cref{sec:results:concealment}). Validating the fence-stripped payload
before composing the retry message would name the underlying violation
in each of these messages. This count bounds how often the change would
alter the feedback, not its effect: we did not run this variant, and
whether a model repairs a violation once it is named is untested, so we
do not claim it would reduce retries or exhaustion.

\paragraph{Evaluation: report composition, not only a rate.}
Over this corpus the attempt-weighted failure rate is largely a statement
about one configuration (\cref{sec:results:rates}), and reading primary
labels alone makes a prompt ablation appear to move failures from
wrapping to length (\cref{sec:results:concealment}). Evaluating a
pipeline change should therefore report the failure composition per
configuration, including violations concealed beneath fences, before
concluding that the change removed a failure.

\paragraph{Will these modes persist?}
A reasonable objection is that providers are engineering these modes
away. Our cross-generation data are descriptive and do not isolate model
generation from other configuration differences, including collection
date, sampling parameters and endpoint. With that caveat, they point in
both directions: \cat{schema\_mirroring} dominates gpt-4o and was not
observed in the GPT-5.6 configurations we tested, while
\cat{markdown\_wrap} was not observed in Sonnet~4.6 and dominates
Sonnet~5 and Haiku~4.5. What differs across these models is which
category dominates. A vocabulary for the categories remains useful as the
distribution moves, which is a weaker claim than predicting that any
particular mode persists.

\section{Threats to Validity}
\label{sec:threats}

We organise threats by the construct, internal, external and conclusion
validity classification of \citet{wohlin2012experimentation}.

\paragraph{Construct validity.}
The taxonomy is an operational diagnostic for the prompt-and-validate
paradigm, not a complete account of structured-output failure. Primary
categories are single-valued, labels follow how a constraint is encoded
rather than what it intends (\cref{sec:setup:taxonomy}), and we quantify
co-occurrence only beneath fences. Human checks of the classifier come from a single labeller who also
wrote it. The development check displayed the classifier's label and is
not independent; the pre-registered held-out check was blind and gives
\pn{heldout.rate} agreement, but covers only the three categories with
attempts left after development, and no second annotator labelled either
sample, so we report no inter-annotator agreement. Three properties limit what this can affect.
Stage-1 categories are syntactic predicates on the raw completion, and
corpus-wide screens that do not depend on any human label find no
missed case among the failures they match (\cref{app:classifier}). Stage-2 categories are a fixed map
from the Pydantic error types stored with each attempt, and rerunning the
released classifier over the database reproduces every stored label.
Disagreements in both checks were resolved against the stored error
record or the raw completion, not the labeller's judgement. The residual
risk lies in the category definitions and the precedence rule rather than
in label noise; we release the annotations and stored completions so that
an independent labeller can test both. We evaluate conformance only, not whether extracted content
is correct. Among attempts whose model-level validators ran, no
cross-field violation occurred (\cref{app:crossfield}), but attempts
rejected earlier, including during JSON parsing, never reached those
checks, so the zero describes a filtered subset of the node-level
validators we implemented. Cross-node referential integrity, whether a
later node's identifiers appear in an earlier node's entities, is not
implemented in our schemas; it is expressible in Pydantic by passing the
earlier output as validation context
(\code{model\_validate\_json(data, context=\ldots)}) and is the natural
next measurement.

\paragraph{Internal validity.}
Repair attempts, which are conditioned on a prior failure, make up a
model-dependent share of each attempt rate; first-attempt rates and run
completion are reported separately for that reason
(\cref{app:breakdowns}). Every retry carries the validation error and no
condition retries without it, so the design does not separate the effect
of the feedback from that of a further attempt. Five models reject an
explicit temperature, and the Anthropic and OpenAI branches use
different token ceilings. Some questions were run more than once; keeping
only the earliest run per question changes any cell's failure rate by at
most \pn{robust.within_cell.max_abs_delta_pp} percentage points
(\cref{app:manifest}). The prompt ablation is exploratory: the variant
cells (\pn{ablation.aggressive.attempts} and \pn{ablation.few_shot.attempts}
attempts against \pn{ablation.baseline.attempts}; \cref{tab:ablation}) were
run after the final baseline batch
rather than interleaved with it and are not paired by question, so
differences between conditions may reflect question composition or
collection time as well as prompt wording. We use it only to show how
concealment changes the reading of failure labels, and draw no conclusion
about prompt effectiveness; a matched, interleaved rerun would be needed
for that.

\paragraph{Which comparisons are matched.}
Across all nine configurations, the HotpotQA cells at the medium tier
share the schema, the baseline system prompt, the retry budget of three
further attempts, and the same
\pn{matched.hotpotqa_medium.shared_questions}-question prefix (\cref{tab:manifest});
these cells are the matched basis for cross-model comparison. The
pooled per-model figures in \cref{tab:models} and \cref{fig:profiles}
also include FinanceBench for six models; per-cell results are in
\cref{tab:domains}. They are not matched in sampling or collection: four
configurations run at \code{temperature=0} and five at provider
defaults, the Anthropic branch sets an output-token ceiling that the
OpenAI branch does not, collection dates range over three weeks
(\cref{tab:provenance}), and the number of runs per question varies by
cell, most for Haiku~4.5 (\cref{tab:manifest}). FinanceBench was
run for six of the nine models, so domain comparisons are within-model
only, and each schema variant was run on one or two models, so
constraint-surface comparisons are within those models. We therefore
describe differences between configurations and do not attribute them to
model capability alone.

\paragraph{External validity.}
The schemas are ours, and one ablation showed that a cap we set was
binding below the output lengths of one model; a different author would
produce a different distribution over categories, and the domain
comparison shows that a different task would too. We cover two vendors by
design, trading breadth for several generations and capability tiers per
vendor, which the within-family comparisons require; a third vendor or an
open-weights model may exhibit modes our taxonomy lacks. We do not
compare against constrained decoding or vendor structured-output modes
(\cref{sec:related}), so we make no claim about how often those
configurations would prevent the failures we report. Questions are fixed
dataset prefixes, the design is not fully crossed, and most failures in
this corpus come from a single model. Most endpoints returned an alias
rather than a pinned version, so later calls to the same identifier may
not reproduce these distributions.

\paragraph{Conclusion validity.}
Attempts are not independent, so all intervals are question-clustered
bootstrap intervals (\cref{sec:setup:basis}) that describe variability
over the observed question set, not over the benchmarks. We do not
interpret cells with a handful of failures, and we treat maximum-gap
divergence statistics as secondary because they are selected to be
extreme (\cref{app:divergence}).
\section{Conclusion}
\label{sec:conclusion}

A validity rate for prompt-enforced structured output does not tell a
practitioner what failed or what a fix will change. Diagnosing a failure
requires keeping both the visible formatting failure and the schema
violation it may conceal; in this corpus, dropping either layer changes
the conclusion an engineer would draw about a repair, a schema change, or
a prompt change. Across
nine model configurations, those with similar rates fail
in different ways (RQ1); the dominant category moves when the binding constraints
change and differs between task domains for the same model (RQ2); and the
most common failure, markdown wrapping, conceals a length or pattern
violation often enough that reading primary labels alone misreads a
prompt ablation and misattributes what a repair loop leaves unrepaired
(RQ3). In both schema interventions we ran, whether a relaxation removed
a failure tracked the sub-cause beneath the label rather than the label
itself. Of the four practices we derive,
three are supported by these data; the fourth, building retry feedback
from the fence-stripped validation, follows from them but is untested. We release the classifier, raw
completions and annotation record so the corpus can be relabelled under
other definitions; the distribution over categories will move as models
change, and the vocabulary is meant to outlast it.

\ifanon\else
\section*{Statements and Declarations}
\label{sec:declarations}

\bmhead{Funding} No funding was received for this work. API usage was
paid for by the author.

\bmhead{Competing interests} The author declares no competing interests.
The author has no financial ties to OpenAI or Anthropic, whose models are
evaluated in this study. The author works as an on-site contractor at
American Express through Data Systems Integration Group; this work was
carried out independently and is unrelated to that role.

\bmhead{Ethics approval} Not applicable. The study involves no human
participants; the datasets used are public.

\bmhead{Consent to participate and consent for publication} Not applicable.

\bmhead{Data availability} The results database, raw completions, the
development and held-out annotations, and the question manifest are
available at \url{https://github.com/amitvermaknw/schema-enforcement} and
archived at \url{https://doi.org/10.5281/zenodo.23586834}.
FinanceBench text is redistributed under CC~BY-NC~4.0 and HotpotQA
questions under CC~BY-SA~4.0 (see \code{DATA\_LICENSES.md}).

\bmhead{Code availability} The harness, classifier and analysis scripts
are available in the same repository.

\bmhead{Use of AI tools}
The author used large language model assistants (Claude) for writing,
editing and coding assistance, and, during classifier development, as an
advisor when interpreting model outputs for the development labels;
every label was assigned by the author. The held-out labels were assigned
by the author without AI assistance on any item; an assistant was
consulted only on the wording of the category definitions. All
experiments were designed, run and verified by the author; every number
reported is generated by scripts from the released results database.

\bmhead{Author contributions} Sole author: conceptualisation,
methodology, software, analysis and writing.
\fi

\begin{appendices}
\section{Cell Manifest, Versions and Sampling}
\label{app:manifest}

\begin{table}[htbp]
\centering\scriptsize
\caption{Run manifest, one row per model, tier, domain and prompt variant.
Runs, attempts and failures use the same prompt-variant filter. \emph{Qs}
is the number of distinct questions; \emph{Collected} is the range of run
timestamps (UTC). Pilot cells (fewer than 30 baseline runs) and
prompt-variant rows are outside the reporting basis.}
\label{tab:manifest}
\resizebox{\linewidth}{!}{\begin{tabular}{lllrrrrrll}
\toprule
Model & Tier & Domain & Prompt & Runs & Qs & Att. & Fail. & Collected & Excluded \\
\midrule
\code{claude-haiku-4-5} & medium & financebench & baseline & 32 & 18 & 263 & 186 & 2026-08-23--2026-08-26 & -- \\
\code{claude-haiku-4-5} & medium & hotpotqa & aggressive & 30 & 30 & 189 & 99 & 2026-08-28--2026-08-28 & prompt\_variant \\
\code{claude-haiku-4-5} & medium & hotpotqa & baseline & 134 & 50 & 1066 & 692 & 2026-08-09--2026-08-28 & -- \\
\code{claude-haiku-4-5} & medium & hotpotqa & few\_shot & 30 & 30 & 176 & 88 & 2026-08-28--2026-08-28 & prompt\_variant \\
\code{claude-opus-4-8} & medium & financebench & baseline & 30 & 30 & 101 & 11 & 2026-08-26--2026-08-26 & -- \\
\code{claude-opus-4-8} & medium & hotpotqa & baseline & 54 & 50 & 164 & 2 & 2026-08-22--2026-08-23 & -- \\
\code{claude-sonnet-4-6} & medium & hotpotqa & baseline & 50 & 50 & 178 & 28 & 2026-08-09--2026-08-09 & -- \\
\code{claude-sonnet-4-6} & medium\_v2 & hotpotqa & baseline & 100 & 50 & 324 & 24 & 2026-08-10--2026-08-11 & -- \\
\code{claude-sonnet-5} & medium & financebench & baseline & 30 & 30 & 114 & 24 & 2026-08-26--2026-08-26 & -- \\
\code{claude-sonnet-5} & medium & hotpotqa & baseline & 52 & 50 & 199 & 43 & 2026-08-22--2026-08-23 & -- \\
\code{gpt-4o} & medium & hotpotqa & baseline & 70 & 50 & 250 & 40 & 2026-08-09--2026-08-09 & -- \\
\code{gpt-4o-mini} & medium & financebench & baseline & 5 & 5 & 28 & 13 & 2026-08-22--2026-08-22 & pilot \\
\code{gpt-4o-mini} & medium & hotpotqa & baseline & 70 & 50 & 242 & 33 & 2026-08-09--2026-08-09 & -- \\
\code{gpt-5.6-luna} & medium & financebench & baseline & 30 & 30 & 97 & 7 & 2026-08-23--2026-08-23 & -- \\
\code{gpt-5.6-luna} & medium & hotpotqa & baseline & 55 & 50 & 169 & 4 & 2026-08-22--2026-08-22 & -- \\
\code{gpt-5.6-luna} & medium\_v3\_loose\_regex & hotpotqa & baseline & 2 & 2 & 6 & 0 & 2026-08-27--2026-08-27 & pilot \\
\code{gpt-5.6-sol} & medium & financebench & baseline & 30 & 30 & 96 & 6 & 2026-08-23--2026-08-23 & -- \\
\code{gpt-5.6-sol} & medium & hotpotqa & baseline & 52 & 50 & 159 & 3 & 2026-08-22--2026-08-22 & -- \\
\code{gpt-5.6-sol} & medium\_v3\_loose\_regex & hotpotqa & baseline & 50 & 50 & 155 & 5 & 2026-08-27--2026-08-27 & -- \\
\code{gpt-5.6-terra} & medium & financebench & baseline & 30 & 30 & 98 & 8 & 2026-08-23--2026-08-23 & -- \\
\code{gpt-5.6-terra} & medium & hotpotqa & baseline & 54 & 50 & 168 & 6 & 2026-08-22--2026-08-22 & -- \\
\code{gpt-5.6-terra} & medium\_v3\_loose\_regex & hotpotqa & baseline & 50 & 50 & 154 & 4 & 2026-08-27--2026-08-27 & -- \\
\bottomrule
\end{tabular}
}
\end{table}

The full corpus of \pn{full.attempts} attempts decomposes into the
\pn{reported.attempts} on the reporting basis, \pn{excluded_pilot.attempts}
in pilot cells below the 30-run threshold, which was fixed before any
results were examined, and \pn{excluded_variant.attempts} from the
\pn{excluded_variant.runs} prompt-variant runs of \cref{tab:ablation}.
Runs, attempts and failures are counted under one filter, so variant runs
appear in their own rows rather than inflating the baseline cell whose
configuration they share.

\paragraph{Sampling.}
HotpotQA questions are the first records of the validation split in the
distractor configuration, in dataset order; FinanceBench questions are the
first records, in dataset order, of the public sample released as
\code{PatronusAI/financebench} on HuggingFace: a subset of the full
FinanceBench question set, distributed as a single \code{train} split. Evidence
is truncated to 2{,}000 characters with a truncation marker inserted. The
datasets are redistributed in the artifact under their licences: FinanceBench under CC~BY-NC~4.0 and HotpotQA under CC~BY-SA~4.0. There is no shuffle and no
seed. Collection ran from \pn{collection.date_min} to
\pn{collection.date_max}; per-cell date ranges and distinct-question
counts are in \cref{tab:manifest}. The Haiku~4.5 FinanceBench cell covers
\pn{c.claude-haiku-4-5@medium@financebench.questions} distinct questions,
fewer than the other FinanceBench cells. Exact question identities, as
dataset record hashes, are in the released \code{question\_manifest.csv}.

\paragraph{Resolved versions and request parameters.}
Only the Haiku request names a dated snapshot, which was returned
unchanged. \code{gpt-4o} resolves to \code{gpt-4o-2024-08-06} and
\code{gpt-4o-mini} to \code{gpt-4o-mini-2024-07-18}; the Anthropic models
and the GPT-5.6 family return the requested alias, which identifies the
endpoint but not an immutable model version, so later calls to the same
identifier may not reproduce these distributions. Version capture was
added after the first sweeps: \pn{versions.gpt-4o.precapture} of
\pn{versions.gpt-4o.attempts} \code{gpt-4o} attempts and
\pn{versions.gpt-4o-mini.precapture} of \pn{versions.gpt-4o-mini.attempts}
\code{gpt-4o-mini} attempts on the reporting basis predate it.
\Cref{tab:provenance} gives the requested and returned identifiers, the
API route, the collection window and the sampling parameters sent by the
client for each model.

\paragraph{Validator and client versions.}
The Pydantic version used during collection was not recorded; the
harness then required version 2.7 or later within major version 2.
Analysis used pydantic 2.13.4 with pydantic-core 2.46.4, which the
released requirements pin. Revalidating every stored completion against
its node schema under these versions reproduces the stored validation
outcome of all \pn{full.attempts} attempts
(\code{scripts/revalidate\_all.py}), so every validity judgement in the
corpus is reproducible under the released environment.
The client SDK versions used during collection were likewise not
recorded; the harness required \code{openai}~$\geq$1.40.0,
\code{anthropic}~$\geq$0.34.0 and \code{langgraph}~$\geq$0.2.0. The
SDKs affect only how requests are sent: the request parameters sent by
the client are given per model in \cref{tab:provenance}, and no validity judgement depends on an
SDK, since validation and classification run on the stored completions.

\begin{table}[htbp]
\centering\scriptsize
\caption{Model provenance. \emph{Returned ID} is the model identifier in
the API response, with attempt counts; \emph{not captured} marks attempts
logged before the response identifier was recorded. Temperature and
\code{max\_tokens} are the values sent by the client.}
\label{tab:provenance}
\resizebox{\linewidth}{!}{\begin{tabular}{l>{\raggedright\arraybackslash}p{4.8cm}llll}
\toprule
Requested ID & Returned ID (attempts) & API & Collected & Temp. & max\_tokens \\
\midrule
\code{claude-\allowbreak{}haiku-\allowbreak{}4-\allowbreak{}5-\allowbreak{}20251001} & \emph{as requested}~(1694) & Anthropic & 2026-08-09--08-28 & 0 & 4096 \\
\code{claude-\allowbreak{}opus-\allowbreak{}4-\allowbreak{}8} & \emph{as requested}~(265) & Anthropic & 2026-08-22--08-26 & omitted & 4096 \\
\code{claude-\allowbreak{}sonnet-\allowbreak{}4-\allowbreak{}6} & \emph{as requested}~(502) & Anthropic & 2026-08-09--08-11 & 0 & 4096 \\
\code{claude-\allowbreak{}sonnet-\allowbreak{}5} & \emph{as requested}~(313) & Anthropic & 2026-08-22--08-26 & omitted & 4096 \\
\code{gpt-\allowbreak{}4o} & gpt-\allowbreak{}4o-\allowbreak{}2024-\allowbreak{}08-\allowbreak{}06~(178) \newline \emph{not captured}~(72) & OpenAI & 2026-08-09 & 0 & not set \\
\code{gpt-\allowbreak{}4o-\allowbreak{}mini} & gpt-\allowbreak{}4o-\allowbreak{}mini-\allowbreak{}2024-\allowbreak{}07-\allowbreak{}18~(201) \newline \emph{not captured}~(69) & OpenAI & 2026-08-09--08-22 & 0 & not set \\
\code{gpt-\allowbreak{}5.6-\allowbreak{}luna} & \emph{as requested}~(272) & OpenAI & 2026-08-22--08-27 & omitted & not set \\
\code{gpt-\allowbreak{}5.6-\allowbreak{}sol} & \emph{as requested}~(410) & OpenAI & 2026-08-22--08-27 & omitted & not set \\
\code{gpt-\allowbreak{}5.6-\allowbreak{}terra} & \emph{as requested}~(420) & OpenAI & 2026-08-22--08-27 & omitted & not set \\
\bottomrule
\end{tabular}
}
\end{table}

\paragraph{Analysis provenance.}
The reported numbers were produced by the analysis scripts at commit
\code{\pn{meta.git_head_short}} from the results database whose content
hash begins \code{\pn{meta.db_sha256_short}}. Intervals are question-clustered
bootstrap intervals with \pn{meta.bootstrap_reps} replicates
(\cref{sec:setup:basis}).

\paragraph{Repeated questions.}
The Haiku~4.5 HotpotQA baseline cell contains \pn{robust.cell.runs} runs
over \pn{robust.cell.questions} distinct questions. Restricting it to the
earliest run of each question by timestamp, ties broken by run identifier,
gives a failure rate of \pn{robust.one_run.rate}
(\pn{robust.one_run.failures} of \pn{robust.one_run.attempts}) against
\pn{robust.cell.rate} (\pn{robust.cell.failures} of
\pn{robust.cell.attempts}) for the full cell. Applied to every cell, the
rule changes no cell's failure rate by more than
\pn{robust.within_cell.max_abs_delta_pp} percentage points. We do not
report a pooled rate under deduplication: it removes mostly Haiku~4.5
attempts, so the pooled change reflects reweighting between models rather
than sensitivity to repetition.

\paragraph{Zero-length completion.}
One attempt (\pn{zero_length.model}, \pn{zero_length.node} node) returned
an empty completion with \pn{zero_length.output_tokens} output tokens,
equal to the configured ceiling. It is classified \cat{parse\_error}. The
stop reason was not logged; the record is consistent with the output
budget being exhausted before any text block was emitted.

\section{Schemas and Prompts}
\label{app:schemas}

The baseline schema, its two variants, the system-prompt variants, the
per-node user prompts, the validation and retry loop (including the retry
message) and the Stage-1 predicates are reproduced below from the released
code by \code{scripts/make\_app\_code.sh}. Listings omit comments and
module and function docstrings; class docstrings are kept because Pydantic
emits them as schema descriptions in the prompt. The listings are
generated from the repository at the commit recorded in
\cref{app:manifest}, so they cannot drift from the code that produced the
results.

\IfFileExists{generated/app_code.tex}{

\paragraph{Baseline schema} (\code{src/schemaeval/schemas/medium.py})
\begin{lstlisting}[language=Python]
import re
from typing import List, Literal

from pydantic import BaseModel, Field, model_validator

CANONICAL_ID_PATTERN = r"^[a-z][a-z0-9_]{2,30}$"
_ID_RE = re.compile(CANONICAL_ID_PATTERN)

class Entity(BaseModel):
    name: str = Field(min_length=1, max_length=100)
    entity_type: Literal["person", "place", "organization", "concept", "other"]
    canonical_id: str = Field(
        pattern=CANONICAL_ID_PATTERN,
        description="lowercase_snake_case identifier, e.g. 'manchester_united'",
    )

class IdentifyOutput(BaseModel):
    """Node 1: entities in the question, with a designated primary."""

    entities: List[Entity] = Field(min_length=2, max_length=5)
    question_type: Literal[
        "who", "what", "when", "where", "why", "how", "comparison"
    ]
    primary_entity_id: str = Field(
        description="Must match one of entities[].canonical_id exactly"
    )
    entity_count: int = Field(
        ge=2, le=5, description="Must equal len(entities)"
    )

    @model_validator(mode="after")
    def _consistency(self):
        ids = {e.canonical_id for e in self.entities}
        if self.primary_entity_id not in ids:
            raise ValueError(
                f"primary_entity_id '{self.primary_entity_id}' not in "
                f"entities canonical_ids {sorted(ids)}"
            )
        if self.entity_count != len(self.entities):
            raise ValueError(
                f"entity_count={self.entity_count} != "
                f"len(entities)={len(self.entities)}"
            )
        return self

class Fact(BaseModel):
    entity_id: str = Field(
        pattern=CANONICAL_ID_PATTERN,
        description="Reference to a canonical_id from the identify step",
    )
    fact: str = Field(min_length=10, max_length=400)
    relevance: float = Field(ge=0.0, le=1.0)

class GatherOutput(BaseModel):
    """Node 2: facts about entities, with consistency count."""

    facts: List[Fact] = Field(min_length=2, max_length=8)
    fact_count: int = Field(
        ge=2, le=8, description="Must equal len(facts)"
    )
    coverage_notes: str = Field(min_length=20, max_length=250)

    @model_validator(mode="after")
    def _consistency(self):
        if self.fact_count != len(self.facts):
            raise ValueError(
                f"fact_count={self.fact_count} != len(facts)={len(self.facts)}"
            )
        return self

class AnswerOutput(BaseModel):
    """Node 3: final answer with confidence and its categorical bucket."""

    answer: str = Field(min_length=20, max_length=400)
    supporting_entity_ids: List[str] = Field(min_length=1, max_length=5)
    confidence: float = Field(ge=0.0, le=1.0)
    confidence_bucket: Literal["low", "medium", "high"] = Field(
        description=(
            "Must match confidence numerically: "
            "confidence < 0.5 => 'low'; "
            "0.5 <= confidence <= 0.8 => 'medium'; "
            "confidence > 0.8 => 'high'"
        )
    )

    @model_validator(mode="after")
    def _consistency(self):
        expected = (
            "low" if self.confidence < 0.5
            else "high" if self.confidence > 0.8
            else "medium"
        )
        if self.confidence_bucket != expected:
            raise ValueError(
                f"confidence={self.confidence} implies bucket='{expected}', "
                f"got '{self.confidence_bucket}'"
            )
        for eid in self.supporting_entity_ids:
            if not _ID_RE.match(eid):
                raise ValueError(
                    f"supporting_entity_id '{eid}' fails canonical_id pattern "
                    f"'{CANONICAL_ID_PATTERN}'"
                )
        return self
\end{lstlisting}

\paragraph{System-prompt variants} (\code{src/schemaeval/prompts.py})
\begin{lstlisting}[language=Python]
import json

_BACKTICK_FENCE = "`" * 3

def _baseline(schema_json_str: str) -> str:
    return (
        "You must respond with a single JSON object matching this schema.\n"
        "Return ONLY the JSON — no markdown code fences, no prose, no explanation.\n\n"
        f"Schema:\n{schema_json_str}"
    )

def _aggressive(schema_json_str: str) -> str:
    return (
        "CRITICAL: You MUST respond with a single JSON object matching the "
        "schema below.\n"
        "Do NOT use markdown formatting.\n"
        f"Do NOT wrap your response in code fences ({_BACKTICK_FENCE}json "
        f"or {_BACKTICK_FENCE}).\n"
        "Do NOT include any prose, explanation, or commentary.\n"
        "Your response MUST begin with { as the very first character.\n"
        "Any response starting with backticks or prose will be rejected.\n\n"
        f"Schema:\n{schema_json_str}"
    )

def _few_shot(schema_json_str: str) -> str:
    return (
        "You must respond with a single JSON object matching this schema.\n"
        "Return ONLY the JSON — no markdown code fences, no prose, no explanation.\n\n"
        "Example of a CORRECT response format:\n"
        '{"example_field": "example_value", "count": 3}\n\n'
        "Example of an INCORRECT response format (do not do this):\n"
        f"{_BACKTICK_FENCE}json\n"
        '{"example_field": "example_value", "count": 3}\n'
        f"{_BACKTICK_FENCE}\n\n"
        "The second example is wrong because it uses markdown code fences. "
        "Do not wrap JSON in backticks or code blocks.\n\n"
        f"Schema:\n{schema_json_str}"
    )

_PROMPT_BUILDERS = {
    "baseline": _baseline,
    "aggressive": _aggressive,
    "few_shot": _few_shot,
}

def build_system_prompt(schema: type, variant: str = "baseline") -> str:
    if variant not in _PROMPT_BUILDERS:
        raise ValueError(
            f"Unknown prompt variant '{variant}'. "
            f"Available: {sorted(_PROMPT_BUILDERS.keys())}"
        )
    schema_json_str = json.dumps(schema.model_json_schema(), indent=2)
    return _PROMPT_BUILDERS[variant](schema_json_str)

def available_variants() -> list[str]:
    return sorted(_PROMPT_BUILDERS.keys())
\end{lstlisting}

\paragraph{Per-node user prompts} (\code{src/schemaeval/topologies/sequential.py})
\begin{lstlisting}[language=Python]
def run_sequential_graph(
    question: str,
    paradigm_fn: ParadigmFn,
    model: str = "gpt-4o-mini",
    max_retries: int = 3,
    schemas: Optional[tuple] = None,
) -> dict:
    if schemas is None:
        schemas = get_schemas("medium")
    IdentifyOutput, GatherOutput, AnswerOutput = schemas

    run_id = str(uuid.uuid4())
    per_node: dict[str, dict] = {}

    prompt1 = (
        f"Identify the key entities and question type in this question:\n"
        f"{question}\n\n"
        "For each entity, provide a canonical_id in lowercase snake_case "
        "(e.g. 'manchester_united', 'kansas_university')."
    )
    identify_out, m1 = paradigm_fn(
        prompt1, IdentifyOutput, model=model, max_retries=max_retries
    )
    per_node["identify"] = m1
    if identify_out is None:
        return _package(run_id, question, per_node, final_answer=None)

    entities_str = "\n".join(
        f"- name='{e.name}' type={e.entity_type} canonical_id={e.canonical_id}"
        for e in identify_out.entities
    )
    prompt2 = (
        f"Question: {question}\n\n"
        f"Entities to research:\n{entities_str}\n\n"
        "For each entity, provide 1-3 facts relevant to answering the question. "
        "Use the exact canonical_id values shown above when referencing entities."
    )
    gather_out, m2 = paradigm_fn(
        prompt2, GatherOutput, model=model, max_retries=max_retries
    )
    per_node["gather"] = m2
    if gather_out is None:
        return _package(run_id, question, per_node, final_answer=None)

    facts_str = "\n".join(
        f"- entity_id={f.entity_id}: {f.fact}" for f in gather_out.facts
    )
    prompt3 = (
        f"Question: {question}\n\n"
        f"Known facts:\n{facts_str}\n\n"
        "Synthesize a final answer. In supporting_entity_ids, list the "
        "canonical_id values (snake_case) of the entities that support your answer."
    )
    answer_out, m3 = paradigm_fn(
        prompt3, AnswerOutput, model=model, max_retries=max_retries
    )
    per_node["answer"] = m3
    final_answer = answer_out.answer if answer_out else None

    return _package(run_id, question, per_node, final_answer)
\end{lstlisting}

\paragraph{Validation and retry loop} (\code{src/schemaeval/paradigms/strict\_repair.py})
\begin{lstlisting}[language=Python]
def call_llm_strict_repair(
    prompt: str,
    schema: type[BaseModel],
    model: str = "gpt-4o-mini",
    max_retries: int = 3,
    temperature: float = 0.0,
    prompt_variant: str = "baseline",
) -> tuple[Optional[BaseModel], dict]:
    system_msg = build_system_prompt(schema, variant=prompt_variant)
    messages = [
        {"role": "system", "content": system_msg},
        {"role": "user", "content": prompt},
    ]

    total_in = 0
    total_out = 0
    num_repairs = 0
    last_error: Optional[str] = None
    attempts: list[dict] = []
    start = time.time()

    for attempt_num in range(max_retries + 1):
        try:
            call = _dispatch(messages, model=model, temperature=temperature)
        except Exception as e:
            last_error = f"api_error: {e}"
            attempts.append({
                "attempt_num": attempt_num,
                "raw_output": None,
                "input_tokens": 0,
                "output_tokens": 0,
                "validation_ok": False,
                "resolved_model_version": None,
                "error_summary": {
                    "categories": ["api_error"],
                    "primary_category": "api_error",
                    "errors": [{
                        "category": "api_error",
                        "loc": "", "type": type(e).__name__, "msg": str(e),
                    }],
                },
            })
            break

        total_in += call["input_tokens"]
        total_out += call["output_tokens"]
        raw = call["raw"]

        error_summary = None
        validated = None
        try:
            validated = schema.model_validate_json(raw)
        except ValidationError as ve:
            error_summary = summarize_validation_error(ve, raw_output=raw)
            last_error = str(ve)
        except (ValueError, json.JSONDecodeError) as pe:
            error_summary = summarize_parse_error(pe, raw_output=raw)
            last_error = str(pe)

        attempts.append({
            "attempt_num": attempt_num,
            "raw_output": raw,
            "input_tokens": call["input_tokens"],
            "output_tokens": call["output_tokens"],
            "validation_ok": validated is not None,
            "resolved_model_version": call.get("resolved_model"),
            "error_summary": error_summary,
        })

        if validated is not None:
            return validated, _metrics(total_in, total_out, num_repairs, start,
                                       success=True, error=None, attempts=attempts)

        num_repairs += 1
        if attempt_num < max_retries:
            messages.append({"role": "assistant", "content": raw})
            messages.append({
                "role": "user",
                "content": (
                    "Your JSON failed validation with these errors:\n"
                    f"{last_error}\n\n"
                    "Return corrected JSON matching the schema. "
                    "ONLY JSON, no prose."
                ),
            })

    return None, _metrics(total_in, total_out, num_repairs, start,
                          success=False, error=last_error, attempts=attempts)
\end{lstlisting}

\paragraph{Stage-1 predicates} (\code{src/schemaeval/errors.py})
\begin{lstlisting}[language=Python]
_SCHEMA_META_KEYS = {
    "description", "properties", "required", "title", "type",
    "additionalProperties", "$defs", "$schema", "$ref",
    "definitions", "items", "anyOf", "allOf", "oneOf",
}

_SCHEMA_META_KEY_THRESHOLD = 2

_EMBEDDED_JSON_RE = re.compile(r"\{.*\}", re.DOTALL)

def _looks_like_schema_mirroring(raw_output: str) -> bool:
    if not raw_output:
        return False
    try:
        parsed = json.loads(raw_output)
    except (json.JSONDecodeError, ValueError):
        stripped = _strip_code_fences(raw_output)
        if stripped == raw_output:
            return False
        try:
            parsed = json.loads(stripped)
        except (json.JSONDecodeError, ValueError):
            return False
    if not isinstance(parsed, dict):
        return False
    root_keys = set(parsed.keys())
    return len(root_keys & _SCHEMA_META_KEYS) >= _SCHEMA_META_KEY_THRESHOLD

def _looks_like_markdown_wrap(raw_output: str) -> bool:
    if not raw_output:
        return False
    stripped_start = raw_output.lstrip()
    return stripped_start.startswith("```")

def _looks_like_prose_preamble(raw_output: str) -> bool:
    if not raw_output:
        return False
    stripped_start = raw_output.lstrip()
    if not stripped_start:
        return False
    first_char = stripped_start[0]
    if first_char in ("{", "["):
        return False
    if stripped_start.startswith("```"):
        return False
    match = _EMBEDDED_JSON_RE.search(raw_output)
    if not match:
        return False
    prose_prefix_len = match.start()
    return prose_prefix_len >= 20

def _strip_code_fences(raw_output: str) -> str:
    stripped = raw_output.strip()
    if stripped.startswith("```"):
        stripped = re.sub(r"^```[a-zA-Z]*\n?", "", stripped, count=1)
        stripped = re.sub(r"\n?```\s*$", "", stripped, count=1)
    return stripped

def _heuristic_category(raw_output: Optional[str]) -> Optional[str]:
    if not raw_output:
        return None
    if _looks_like_markdown_wrap(raw_output):
        return "markdown_wrap"
    if _looks_like_prose_preamble(raw_output):
        return "prose_preamble"
    if _looks_like_schema_mirroring(raw_output):
        return "schema_mirroring"
    return None
\end{lstlisting}

\paragraph{medium\_v2 changes}
\begin{lstlisting}
--- src/schemaeval/schemas/medium.py
+++ src/schemaeval/schemas/medium_v2.py
@@ -16,6 +16,4 @@
 
 class IdentifyOutput(BaseModel):
-    """Node 1: entities in the question, with a designated primary."""
-
     entities: List[Entity] = Field(min_length=2, max_length=5)
     question_type: Literal[
@@ -49,15 +47,13 @@
         description="Reference to a canonical_id from the identify step",
     )
-    fact: str = Field(min_length=10, max_length=400)
+    fact: str = Field(min_length=10, max_length=800)      # was 400
     relevance: float = Field(ge=0.0, le=1.0)
 
 class GatherOutput(BaseModel):
-    """Node 2: facts about entities, with consistency count."""
-
     facts: List[Fact] = Field(min_length=2, max_length=8)
     fact_count: int = Field(
         ge=2, le=8, description="Must equal len(facts)"
     )
-    coverage_notes: str = Field(min_length=20, max_length=250)
+    coverage_notes: str = Field(min_length=20, max_length=1000)  # was 250
 
     @model_validator(mode="after")
@@ -70,7 +66,5 @@
 
 class AnswerOutput(BaseModel):
-    """Node 3: final answer with confidence and its categorical bucket."""
-
-    answer: str = Field(min_length=20, max_length=400)
+    answer: str = Field(min_length=20, max_length=1200)   # was 400
     supporting_entity_ids: List[str] = Field(min_length=1, max_length=5)
     confidence: float = Field(ge=0.0, le=1.0)
\end{lstlisting}

\paragraph{medium\_v3 changes}
\begin{lstlisting}
--- src/schemaeval/schemas/medium.py
+++ src/schemaeval/schemas/medium_v3_loose_regex.py
@@ -4,5 +4,5 @@
 from pydantic import BaseModel, Field, model_validator
 
-CANONICAL_ID_PATTERN = r"^[a-z][a-z0-9_]{2,30}$"
+CANONICAL_ID_PATTERN = r"^[a-zA-Z0-9][a-zA-Z0-9_\-]{2,30}$"
 _ID_RE = re.compile(CANONICAL_ID_PATTERN)
 
@@ -12,5 +12,5 @@
     canonical_id: str = Field(
         pattern=CANONICAL_ID_PATTERN,
-        description="lowercase_snake_case identifier, e.g. 'manchester_united'",
+        description="alphanumeric identifier, may include hyphens/underscores",
     )
 
\end{lstlisting}
}{\emph{[Run \code{review3.py appcode} to generate \code{generated/app\_code.tex}.]}}

The medium\_v2 variant changes the three character caps and omits the
three class docstrings; the medium\_v3\_loose\_regex variant changes
\code{CANONICAL\_ID\_PATTERN} and the \code{canonical\_id} description.
The diffs above show every change, including those to prompt-visible
text.

\section{Classifier Validation}
\label{app:classifier}

\paragraph{Provenance.}
Development labelling was performed against an earlier classifier in which
Pydantic error-type mapping preceded the behavioural predicates.
Schema-mirrored outputs were consequently labelled \cat{missing\_field}
and prose-preamble outputs \cat{parse\_error}. We corrected the order and
reclassified the full corpus from stored completions and error details,
without new model calls. Running the released \code{reclassify\_all.py},
which applies both stages of \code{errors.py}, in dry-run mode over the
released database reproduces the stored primary label of all
\pn{full.failures} failed attempts (zero changes).

\paragraph{Agreement.}
The sample was drawn from all \pn{full.failures} failures in the corpus,
including pilot cells and prompt variants. \Cref{tab:validation} reports
precision and recall of the corrected classifier against the author's
unadjudicated development labels, assigned with the classifier's label
displayed. The four disagreements were adjudicated as labelling
errors: three outputs labelled \cat{enum\_violation} carried a
\code{string\_pattern\_mismatch} on an identifier, and one labelled \cat{markdown\_wrap} contained no fence. The first three
are checkable against the stored error record and the fourth against the raw
completion, without reference to the labeller. We headline unadjudicated agreement
because adjudication uses evidence the labeller did not consult. The sample
contains every \cat{enum\_violation}, \cat{prose\_preamble} and
\cat{schema\_mirroring} failure in the corpus, so a held-out check with a
frozen classifier can test only the remaining categories.

\paragraph{Held-out check.}
The protocol, sample and seed were committed to the repository before
labelling, in \code{PREREGISTRATION.md} under \code{results/heldout/}.
The labelling screen showed model, dataset, tier, node and the raw output, truncated in the middle
for long outputs as in the development round, but not the classifier's
label, and offered no default answer. Labelling was restarted once, after
the precedence rule was reread and before any score was computed; the
restarted pass is the one reported. Agreement was
\pn{heldout.cat.markdown_wrap.agree} of \pn{heldout.cat.markdown_wrap.n}
for \cat{markdown\_wrap}, \pn{heldout.cat.regex_pattern.agree} of
\pn{heldout.cat.regex_pattern.n} for \cat{regex\_pattern}, and
\pn{heldout.cat.length_bound.agree} of \pn{heldout.cat.length_bound.n}
for \cat{length\_bound}. The annotator marked \pn{heldout.ambiguous}
items ambiguous; excluding them, agreement is
\pn{heldout.nonambiguous.rate} (post hoc). The per-item disagreement
record is released as \code{results/heldout/adjudication.md}.

\begin{table}[ht]
\centering\small
\caption{Corrected classifier against unadjudicated development labels
($n=\pn{validation.n}$), full corpus frame. Recall is computed within the
label-stratified sample, not over the corpus.}
\label{tab:validation}
\adjustbox{max width=\textwidth}{\begin{tabular}{@{}lrll@{}}
\toprule
Category & Predicted & Precision & Recall \\
\midrule
\cat{markdown\_wrap} & 25 & 25/25 & 25/26 \\
\cat{length\_bound} & 30 & 30/30 & 30/30 \\
\cat{regex\_pattern} & 30 & 27/30 & 27/27 \\
\cat{schema\_mirroring} & 42 & 42/42 & 42/42 \\
\cat{prose\_preamble} & 9 & 8/9 & 8/8 \\
\cat{enum\_violation} & 8 & 8/8 & 8/11 \\
\bottomrule
\end{tabular}
}
\end{table}

\paragraph{Corpus screens.}
Because the sample is stratified by predicted label, it cannot establish
corpus recall. For the Stage-1 categories we screened every failure in the full corpus
with relaxed screens for wrapping and preambles, and a keyword-restricted
screen for mirroring. A fence at
any position matches \pn{screen.fence_any.matches} failures, of which
\pn{screen.fence_any.markdown_wrap} are classified \cat{markdown\_wrap}
and \pn{screen.fence_any.prose_preamble} \cat{prose\_preamble} (prose
before a later fence). A \code{"properties"} key matches
\pn{screen.properties.matches} failures, of which
\pn{screen.properties.schema_mirroring} are classified
\cat{schema\_mirroring}; this screen tests one meta-keyword where the
detector requires two, so it checks for missed mirroring cases only among
failures containing that key. An opening character other than a brace, bracket or fence
matches \pn{screen.nonjson_opener.matches} non-empty failures, of which
\pn{screen.nonjson_opener.prose_preamble} are \cat{prose\_preamble}.

\section{Identifier Perturbation}
\label{app:perturbation}

\begin{table}[ht]
\centering\small
\caption{Sub-causes of violating identifiers in \cat{regex\_pattern}
failures, gpt-5.6-sol and gpt-5.6-terra, HotpotQA. Counts are
identifiers, not attempts.}
\label{tab:perturbation}
\adjustbox{max width=\textwidth}{\begin{tabular}{@{}lrr@{}}
\toprule
Sub-cause & Baseline pattern & Loosened pattern \\
\midrule
Leading digit & 4 & 0 \\
Length overflow & 4 & 10 \\
Unicode contamination & 0 & 1 \\
Other character & 0 & 0 \\
\midrule
Violating identifiers & 8 & 11 \\
\bottomrule
\end{tabular}
}
\end{table}

The baseline tier uses \idpattern{} and the perturbed tier \idpatternloose{};
both retain the 31-character ceiling, and the loosened pattern also
governs the supporting-identifier check inside the answer validator.
Sub-causes are assigned mechanically per identifier: non-ASCII characters,
a leading digit not admitted by the first-character class, other
characters outside the class, and length outside 3--31. The cells hold
\pn{pert.medium.gpt-5.6-sol.failures} and
\pn{pert.medium.gpt-5.6-terra.failures} failures under the baseline
pattern and \pn{pert.medium_v3_loose_regex.gpt-5.6-sol.failures} and
\pn{pert.medium_v3_loose_regex.gpt-5.6-terra.failures} under the loosened
pattern, too few failures for precise comparative inference. Pooled over
the four cells, \pn{pert.pooled.regex_failures} of
\pn{pert.pooled.failures} failing attempts were assigned
\cat{regex\_pattern}, and these contained \pn{pert.pooled.identifiers}
violating identifiers. Sub-causes are annotated as multilabel rather than
by precedence, and no identifier carried more than one; non-ASCII
characters are counted as \emph{unicode} and excluded from the
invalid-character check. The single Unicode case under the loosened
pattern is an otherwise lowercase band name with a non-Latin combining
mark inserted.

\section{Per-Domain and First-Attempt Breakdowns}
\label{app:breakdowns}

\begin{table}[htbp]
\centering\scriptsize
\caption{Per-cell results, reporting basis.}
\label{tab:domains}
\adjustbox{max width=\textwidth}{\begin{tabular}{@{}lllrrrl@{}}
\toprule
Model & Tier & Domain & Att. & Rate & Att./run & Dominant \\
\midrule
\code{claude-haiku-4-5} & medium & financebench & 263 & 70.7\% & 8.22 & \cat{markdown\_wrap} \\
\code{claude-haiku-4-5} & medium & hotpotqa & 1{,}066 & 64.9\% & 7.96 & \cat{markdown\_wrap} \\
\code{claude-opus-4-8} & medium & financebench & 101 & 10.9\% & 3.37 & \cat{length\_bound} \\
\code{claude-opus-4-8} & medium & hotpotqa & 164 & 1.2\% & 3.04 & \cat{regex\_pattern} \\
\code{claude-sonnet-4-6} & medium & hotpotqa & 178 & 15.7\% & 3.56 & \cat{length\_bound} \\
\code{claude-sonnet-4-6} & medium\_v2 & hotpotqa & 324 & 7.4\% & 3.24 & \cat{regex\_pattern} \\
\code{claude-sonnet-5} & medium & financebench & 114 & 21.1\% & 3.80 & \cat{length\_bound} \\
\code{claude-sonnet-5} & medium & hotpotqa & 199 & 21.6\% & 3.83 & \cat{markdown\_wrap} \\
\code{gpt-4o} & medium & hotpotqa & 250 & 16.0\% & 3.57 & \cat{schema\_mirroring} \\
\code{gpt-4o-mini} & medium & hotpotqa & 242 & 13.6\% & 3.46 & \cat{regex\_pattern} \\
\code{gpt-5.6-luna} & medium & financebench & 97 & 7.2\% & 3.23 & \cat{regex\_pattern} \\
\code{gpt-5.6-luna} & medium & hotpotqa & 169 & 2.4\% & 3.07 & \cat{regex\_pattern} \\
\code{gpt-5.6-sol} & medium & financebench & 96 & 6.2\% & 3.20 & \cat{regex\_pattern} \\
\code{gpt-5.6-sol} & medium & hotpotqa & 159 & 1.9\% & 3.06 & \cat{regex\_pattern} \\
\code{gpt-5.6-sol} & medium\_v3\_loose\_regex & hotpotqa & 155 & 3.2\% & 3.10 & \cat{regex\_pattern} \\
\code{gpt-5.6-terra} & medium & financebench & 98 & 8.2\% & 3.27 & \cat{regex\_pattern} \\
\code{gpt-5.6-terra} & medium & hotpotqa & 168 & 3.6\% & 3.11 & \cat{regex\_pattern} \\
\code{gpt-5.6-terra} & medium\_v3\_loose\_regex & hotpotqa & 154 & 2.6\% & 3.08 & \cat{regex\_pattern} \\
\bottomrule
\end{tabular}
}
\end{table}

\begin{table}[htbp]
\centering\scriptsize
\caption{Run completion and offline residual invalidity, reporting basis.
\emph{Att./compl.} is the total number of attempts in the cell,
including those of incomplete runs, divided by the number of completed
runs: the attempt cost per completed run, not the mean number of attempts
a completed run used. \emph{Residual} counts
pure-formatting \cat{markdown\_wrap} failures as recovered; it is computed
on stored attempts and is not the outcome of rerunning the pipeline with
fence removal enabled.}
\label{tab:operational}
\resizebox{\linewidth}{!}{\begin{tabular}{llrrrrrrl}
\toprule
Model & Tier & Runs & Compl. & Compl. \% & Att./compl. & Strict fail \% & Residual \% & 95\% CI \\
\midrule
\code{claude-haiku-4-5} & medium & 166 & 142 & 85.5 & 9.36 & 66.1 & 44.2 & [40.6, 48.0] \\
\code{claude-sonnet-5} & medium & 82 & 82 & 100.0 & 3.82 & 21.4 & 9.6 & [6.3, 13.0] \\
\code{gpt-4o} & medium & 70 & 70 & 100.0 & 3.57 & 16.0 & 16.0 & [11.9, 20.0] \\
\code{claude-sonnet-4-6} & medium & 50 & 50 & 100.0 & 3.56 & 15.7 & 15.7 & [10.2, 21.5] \\
\code{gpt-4o-mini} & medium & 70 & 69 & 98.6 & 3.51 & 13.6 & 13.6 & [8.0, 19.7] \\
\code{claude-sonnet-4-6} & medium\_v2 & 100 & 100 & 100.0 & 3.24 & 7.4 & 7.4 & [3.8, 10.7] \\
\code{gpt-5.6-terra} & medium & 84 & 84 & 100.0 & 3.17 & 5.3 & 5.3 & [3.1, 7.7] \\
\code{claude-opus-4-8} & medium & 84 & 84 & 100.0 & 3.15 & 4.9 & 4.9 & [2.7, 7.4] \\
\code{gpt-5.6-luna} & medium & 85 & 85 & 100.0 & 3.13 & 4.1 & 4.1 & [2.2, 6.3] \\
\code{gpt-5.6-sol} & medium & 82 & 82 & 100.0 & 3.11 & 3.5 & 3.5 & [1.6, 5.7] \\
\code{gpt-5.6-sol} & medium\_v3\_loose\_regex & 50 & 50 & 100.0 & 3.10 & 3.2 & 3.2 & [0.7, 6.2] \\
\code{gpt-5.6-terra} & medium\_v3\_loose\_regex & 50 & 50 & 100.0 & 3.08 & 2.6 & 2.6 & [0.7, 5.1] \\
\emph{Pooled} &  & 973 & 948 & 97.4 & 4.22 & 28.2 & 20.0 & [17.6, 22.5] \\
\bottomrule
\end{tabular}
}
\end{table}

Run completion and attempt failure diverge most where the repair loop does
the most work: Haiku~4.5 fails \pn{g.claude-haiku-4-5@medium.rate} of
attempts yet completes \pn{g.claude-haiku-4-5@medium.completion_rate} of
runs, at a cost of
\pn{g.claude-haiku-4-5@medium.attempts_per_completed_run} attempts per
completed run against \pn{pooled.reported.attempts_per_completed_run}
pooled, where the cost counts every attempt in the cell, including those
of incomplete runs.

\begin{table}[htbp]
\centering\scriptsize
\caption{First attempts by node: valid first attempts over run-nodes
reaching the node, the same after fence removal, and the composition of
failed first attempts. The denominator depends on upstream success, so the
identify node is the only unconditioned comparison. \emph{Retry} rows give
failed first attempts that a later attempt repaired.}
\label{tab:firstattempt}
\resizebox{\linewidth}{!}{\begin{tabular}{lllrrp{5cm}}
\toprule
Model & Tier & Node & Valid/reached & After fence & First-attempt failure composition \\
\midrule
\code{claude-haiku-4-5} & medium & identify & 0/166 & 144/166 & markdown\_wrap 166 \\
\code{claude-haiku-4-5} & medium & gather & 0/166 & 21/166 & markdown\_wrap 166 \\
\code{claude-haiku-4-5} & medium & answer & 0/143 & 96/143 & markdown\_wrap 143 \\
\code{claude-haiku-4-5} & medium & \emph{retry} & 451/475 &  & \emph{failed first attempts later valid} \\
\code{claude-opus-4-8} & medium & identify & 81/84 & 81/84 & regex\_pattern 3 \\
\code{claude-opus-4-8} & medium & gather & 84/84 & 84/84 & -- \\
\code{claude-opus-4-8} & medium & answer & 75/84 & 75/84 & length\_bound 9 \\
\code{claude-opus-4-8} & medium & \emph{retry} & 12/12 &  & \emph{failed first attempts later valid} \\
\code{claude-sonnet-4-6} & medium & identify & 41/50 & 41/50 & regex\_pattern 7, enum\_violation 2 \\
\code{claude-sonnet-4-6} & medium & gather & 41/50 & 41/50 & length\_bound 9 \\
\code{claude-sonnet-4-6} & medium & answer & 42/50 & 42/50 & length\_bound 7, prose\_preamble 1 \\
\code{claude-sonnet-4-6} & medium & \emph{retry} & 26/26 &  & \emph{failed first attempts later valid} \\
\code{claude-sonnet-4-6} & medium\_v2 & identify & 84/100 & 84/100 & regex\_pattern 14, enum\_violation 2 \\
\code{claude-sonnet-4-6} & medium\_v2 & gather & 96/100 & 96/100 & prose\_preamble 4 \\
\code{claude-sonnet-4-6} & medium\_v2 & answer & 96/100 & 96/100 & prose\_preamble 4 \\
\code{claude-sonnet-4-6} & medium\_v2 & \emph{retry} & 24/24 &  & \emph{failed first attempts later valid} \\
\code{claude-sonnet-5} & medium & identify & 53/82 & 69/82 & markdown\_wrap 20, regex\_pattern 9 \\
\code{claude-sonnet-5} & medium & gather & 63/82 & 80/82 & markdown\_wrap 17, length\_bound 1, parse\_error 1 \\
\code{claude-sonnet-5} & medium & answer & 71/82 & 71/82 & length\_bound 11 \\
\code{claude-sonnet-5} & medium & \emph{retry} & 59/59 &  & \emph{failed first attempts later valid} \\
\code{gpt-4o} & medium & identify & 59/70 & 59/70 & regex\_pattern 8, enum\_violation 3 \\
\code{gpt-4o} & medium & gather & 70/70 & 70/70 & -- \\
\code{gpt-4o} & medium & answer & 41/70 & 41/70 & schema\_mirroring 29 \\
\code{gpt-4o} & medium & \emph{retry} & 40/40 &  & \emph{failed first attempts later valid} \\
\code{gpt-4o-mini} & medium & identify & 55/70 & 55/70 & regex\_pattern 15 \\
\code{gpt-4o-mini} & medium & gather & 70/70 & 70/70 & -- \\
\code{gpt-4o-mini} & medium & answer & 60/70 & 60/70 & schema\_mirroring 8, length\_bound 2 \\
\code{gpt-4o-mini} & medium & \emph{retry} & 24/25 &  & \emph{failed first attempts later valid} \\
\code{gpt-5.6-luna} & medium & identify & 77/85 & 77/85 & regex\_pattern 8 \\
\code{gpt-5.6-luna} & medium & gather & 85/85 & 85/85 & -- \\
\code{gpt-5.6-luna} & medium & answer & 82/85 & 82/85 & length\_bound 3 \\
\code{gpt-5.6-luna} & medium & \emph{retry} & 11/11 &  & \emph{failed first attempts later valid} \\
\code{gpt-5.6-sol} & medium & identify & 74/82 & 74/82 & regex\_pattern 8 \\
\code{gpt-5.6-sol} & medium & gather & 82/82 & 82/82 & -- \\
\code{gpt-5.6-sol} & medium & answer & 81/82 & 81/82 & length\_bound 1 \\
\code{gpt-5.6-sol} & medium & \emph{retry} & 9/9 &  & \emph{failed first attempts later valid} \\
\code{gpt-5.6-sol} & medium\_v3\_loose\_regex & identify & 46/50 & 46/50 & regex\_pattern 4 \\
\code{gpt-5.6-sol} & medium\_v3\_loose\_regex & gather & 50/50 & 50/50 & -- \\
\code{gpt-5.6-sol} & medium\_v3\_loose\_regex & answer & 49/50 & 49/50 & regex\_pattern 1 \\
\code{gpt-5.6-sol} & medium\_v3\_loose\_regex & \emph{retry} & 5/5 &  & \emph{failed first attempts later valid} \\
\code{gpt-5.6-terra} & medium & identify & 71/84 & 71/84 & regex\_pattern 12, enum\_violation 1 \\
\code{gpt-5.6-terra} & medium & gather & 84/84 & 84/84 & -- \\
\code{gpt-5.6-terra} & medium & answer & 83/84 & 83/84 & length\_bound 1 \\
\code{gpt-5.6-terra} & medium & \emph{retry} & 14/14 &  & \emph{failed first attempts later valid} \\
\code{gpt-5.6-terra} & medium\_v3\_loose\_regex & identify & 46/50 & 46/50 & regex\_pattern 4 \\
\code{gpt-5.6-terra} & medium\_v3\_loose\_regex & gather & 50/50 & 50/50 & -- \\
\code{gpt-5.6-terra} & medium\_v3\_loose\_regex & answer & 50/50 & 50/50 & -- \\
\code{gpt-5.6-terra} & medium\_v3\_loose\_regex & \emph{retry} & 4/4 &  & \emph{failed first attempts later valid} \\
\bottomrule
\end{tabular}
}
\end{table}

Pooled over the reporting basis, \pn{pooled.reported.fa.valid} of
\pn{pooled.reported.fa.reached} first attempts validated
(\pn{pooled.reported.fa.valid_rate}), rising to
\pn{pooled.reported.fa.valid_after_fence_rate} if pure-formatting wrapped
failures are counted as valid, and \pn{pooled.reported.fa.retry_success}
of the \pn{pooled.reported.fa.failed} failed first attempts were repaired
by a later attempt.

\paragraph{Audit of first-attempt extraction.}
Because every Haiku~4.5 first attempt failed, we checked the extraction
directly against the database. Every run-node has exactly one attempt
numbered zero; attempt numbers are contiguous from zero; no attempt is
logged after a valid attempt in the same run-node; and the validation flag
agrees with the stored category and error record for every attempt in the
corpus. First-attempt validity is non-zero for every other model under the
same query. All \pn{g.claude-haiku-4-5@medium.fa.reached} Haiku first
attempts begin with a code fence in the raw completion, a check independent
of the classifier, and this holds in each collection batch and in both
domains. After the first validation error is returned, fences persist in
\pn{g.claude-haiku-4-5@medium.att1.cat.markdown_wrap} of
\pn{g.claude-haiku-4-5@medium.att1.n} second attempts and in none
thereafter, while \cat{length\_bound} accounts for every failure from the
third attempt onward.

\begin{table}[htbp]
\centering\scriptsize
\caption{Dominant category and fence-strip outcome by cell.
\emph{Pure/wrapped} is pure-formatting wrapped failures over all wrapped
failures. Dominant shares for cells with few failures are descriptive
only.}
\label{tab:domain}
\resizebox{\linewidth}{!}{\begin{tabular}{lllrlrrr}
\toprule
Model & Tier & Domain & Rate \% & Dominant & Share \% & Pure/wrapped & Compound \\
\midrule
\code{claude-haiku-4-5} & medium & financebench & 70.7 & markdown\_wrap & 63.4 & 34/118 & 84 \\
\code{claude-haiku-4-5} & medium & hotpotqa & 64.9 & markdown\_wrap & 75.4 & 256/522 & 266 \\
\code{claude-opus-4-8} & medium & financebench & 10.9 & length\_bound & 90.9 & -- & -- \\
\code{claude-opus-4-8} & medium & hotpotqa & 1.2 & regex\_pattern & 100.0 & -- & -- \\
\code{claude-sonnet-4-6} & medium & hotpotqa & 15.7 & length\_bound & 64.3 & -- & -- \\
\code{claude-sonnet-4-6} & medium\_v2 & hotpotqa & 7.4 & regex\_pattern & 58.3 & -- & -- \\
\code{claude-sonnet-5} & medium & financebench & 21.1 & length\_bound & 41.7 & 5/6 & 1 \\
\code{claude-sonnet-5} & medium & hotpotqa & 21.6 & markdown\_wrap & 81.4 & 32/35 & 3 \\
\code{gpt-4o} & medium & hotpotqa & 16.0 & schema\_mirroring & 72.5 & -- & -- \\
\code{gpt-4o-mini} & medium & hotpotqa & 13.6 & regex\_pattern & 66.7 & -- & -- \\
\code{gpt-5.6-luna} & medium & financebench & 7.2 & regex\_pattern & 57.1 & -- & -- \\
\code{gpt-5.6-luna} & medium & hotpotqa & 2.4 & regex\_pattern & 100.0 & -- & -- \\
\code{gpt-5.6-sol} & medium & financebench & 6.2 & regex\_pattern & 83.3 & -- & -- \\
\code{gpt-5.6-sol} & medium & hotpotqa & 1.9 & regex\_pattern & 100.0 & -- & -- \\
\code{gpt-5.6-sol} & medium\_v3\_loose\_regex & hotpotqa & 3.2 & regex\_pattern & 100.0 & -- & -- \\
\code{gpt-5.6-terra} & medium & financebench & 8.2 & regex\_pattern & 87.5 & -- & -- \\
\code{gpt-5.6-terra} & medium & hotpotqa & 3.6 & regex\_pattern & 83.3 & -- & -- \\
\code{gpt-5.6-terra} & medium\_v3\_loose\_regex & hotpotqa & 2.6 & regex\_pattern & 100.0 & -- & -- \\
\bottomrule
\end{tabular}
}
\end{table}

\section{Model-Level Constraints}
\label{app:crossfield}

Model validators run only after every field-level constraint passes.
On the reporting basis, \pn{crossfield.reached} attempts reached them:
\pn{crossfield.validated} that validated and \pn{modellevel.violations}
that passed field validation and failed the per-item
supporting-identifier check in the answer validator. That check follows
the confidence-bucket check in the same method, so those attempts also
passed it. No attempt matched any of the four cross-field messages
(\pn{crossfield.violations} violations). The remaining
\pn{crossfield.not_reached} failures did not reach the validators; many
failed during JSON parsing. Each node tests only its own checks: two
cross-field checks in identify, one in gather, and one cross-field plus
one per-item pattern check in answer. \pn{hidden.exhausted_attempts} of the
\pn{hidden.n} are the final attempts of one gpt-4o-mini answer node, which
exhausted its retries although each returned error named the required
pattern; it is that model's only incomplete run. Excluding all
\pn{hidden.n} changes no cell's or group's dominant category; gpt-4o-mini's
\cat{regex\_pattern} share moves from
\pn{g.gpt-4o-mini@medium.dominant_share} to
\pn{hidden.gpt4omini.share_without}.

\paragraph{Prompt visibility.}
The system prompt embeds each node's \code{model\_json\_schema()}
(\cref{app:schemas}), so a constraint is visible on the first attempt
if it is a JSON Schema keyword or is stated in a field or model
description. \Cref{tab:visibility} lists every constraint the medium
tier enforces. The only enforced constraint absent from the prompt is
the per-item identifier pattern on \code{supporting\_entity\_ids}, the
check behind the \pn{modellevel.violations} model-level failures above;
the four cross-field checks, each stated in a description, produced
\pn{crossfield.violations} violations. Conversely, the description of
\code{Fact.entity\_id} asks for a reference to an identify-step
identifier, but only its pattern is enforced.

\begin{table}[ht]
\centering
\small
\caption{Prompt visibility of the constraints enforced by the medium
tier. The prompt carries each node's JSON Schema, including field and
model descriptions.}
\label{tab:visibility}
\begin{adjustbox}{max width=\textwidth}
\begin{tabular}{@{}lll@{}}
\toprule
Constraint & Enforced by & Visible in prompt as \\
\midrule
Character caps, list-length bounds, numeric ranges & field & schema keyword \\
Enumerations (entity type, question type, bucket) & field & schema keyword \\
Identifier pattern on \code{canonical\_id}, \code{entity\_id} & field & keyword and description \\
\code{primary\_entity\_id} among entity identifiers & validator & description \\
\code{entity\_count}, \code{fact\_count} equal list lengths & validator & description \\
Bucket coherent with \code{confidence} & validator & description with thresholds \\
Identifier pattern on each \code{supporting\_entity\_ids} item & validator & \textbf{not visible} \\
\bottomrule
\end{tabular}
\end{adjustbox}
\end{table}

\section{Divergence Metric}
\label{app:divergence}

For each pair of model--tier groups and each category we take the
absolute difference in the share of the group's failures held by that
category. Groups pool task domains and use the reporting basis; of
\pn{divergence.groups_total} groups with any failure,
\pn{divergence.groups_qualifying} have at least 20 failures, giving
\pn{divergence.pairs} pairs. The two largest values arise on different
categories from opposite vendors, \pn{divergence.top.0.pp} and
\pn{divergence.runner_up_category_pp} points, and differ by
\pn{divergence.gap_pp} points, so we attach no significance to their
order. The maximum of \pn{divergence.top.0.pp} points is itself attained
by \pn{divergence.top.n_tied} pairs, each setting Haiku~4.5 against a
group with no \cat{markdown\_wrap} failures. Without the threshold the
maximum is \pn{divergence.unrestricted.max.pp} points in
\cat{regex\_pattern}, attained by \pn{divergence.unrestricted.max.n_tied}
pairs, Haiku~4.5 at medium against each of gpt-5.6-sol and gpt-5.6-terra
at medium\_v3, whose failures are all \cat{regex\_pattern}; those groups
have very few failures. Maximum-gap statistics are selected to be extreme
and do not summarise the distributions; we report them as secondary.

\begin{table}[ht]
\centering\small
\caption{Largest category-share differences among qualifying groups.}
\adjustbox{max width=\textwidth}{\begin{tabular}{@{}lllr@{}}
\toprule
Category & Group A & Group B & pp \\
\midrule
markdown\_wrap & \code{claude-haiku-4-5@medium} & \code{claude-sonnet-4-6@medium} & 72.9 \\
markdown\_wrap & \code{claude-haiku-4-5@medium} & \code{claude-sonnet-4-6@medium\_v2} & 72.9 \\
markdown\_wrap & \code{claude-haiku-4-5@medium} & \code{gpt-4o@medium} & 72.9 \\
markdown\_wrap & \code{claude-haiku-4-5@medium} & \code{gpt-4o-mini@medium} & 72.9 \\
schema\_mirroring & \code{claude-haiku-4-5@medium} & \code{gpt-4o@medium} & 72.5 \\
schema\_mirroring & \code{claude-sonnet-4-6@medium} & \code{gpt-4o@medium} & 72.5 \\
schema\_mirroring & \code{claude-sonnet-4-6@medium\_v2} & \code{gpt-4o@medium} & 72.5 \\
schema\_mirroring & \code{claude-sonnet-5@medium} & \code{gpt-4o@medium} & 72.5 \\
\bottomrule
\end{tabular}
}
\end{table}
\end{appendices}

\bibliography{references}

\end{document}